\pdfoutput=1
\documentclass[preprint,12pt]{elsarticle}

\usepackage{amssymb}
\usepackage{graphicx}
\usepackage{subfigure}
\usepackage{caption}
\usepackage{hyperref}
\usepackage{siunitx}
\usepackage{psfrag}
\usepackage{pstool}
\usepackage{changes}
\usepackage{commath}
\usepackage[normalem]{ulem}
\usepackage[numbers]{natbib}
\usepackage{floatrow}
\usepackage{placeins}
\usepackage[symbol]{footmisc}
\usepackage{bm}
\usepackage{amsmath}
\usepackage{chngcntr}

\begin{document}

\begin{frontmatter}



\title{An in-situ synchrotron diffraction study of stress relaxation in titanium: Effect of temperature and oxygen on cold dwell fatigue}

%\def\correspondingauthor{\footnote{*Corresponding author.}}
\author[add1]{Yi Xiong\corref{*}}
\cortext[*]{Corresponding author}
\address[add1]{Department of Materials, University of Oxford, Parks Road, Oxford, OX1 3PH, United Kingdom}
\ead{yi.xiong@materials.ox.ac.uk}

\author[add1]{Phani S. Karamched}

\author[add2]{Chi-Toan Nguyen}
\address[add2]{Safran SA, Safran Tech, Department of Materials and Processes, 78772 Many-les-Hameaux, France}

\author[add3]{David M. Collins}
\address[add3]{School of Metallurgy and Materials, University of Birmingham, Edgbaston, Birmingham, B15 2TT, United Kingdom}

\author[add4]{Nicol\`{o} Grilli}
\address[add4]{Department of Engineering Science, University of Oxford, Parks Road, Oxford, OX1 3PJ, United Kingdom}
\author[add1]{Christopher M. Magazzeni}
\author[add4,add1]{Edmund Tarleton}
\author[add1]{Angus J. Wilkinson}

\begin{abstract}
There is a long-standing technological problem in which a stress dwell during cyclic loading at room temperature in Ti causes a significant fatigue life reduction. It is thought that localised time dependent plasticity in ‘soft’ grains favourably oriented for easy plastic slip leads to load shedding and an increase in stress within a neighbouring ‘hard’ grain poorly oriented for easy slip. Quantifying this time dependent plasticity process is key to understand the complex cold dwell fatigue problem. Knowing the effect of operating temperature and oxygen content on cold dwell fatigue will be beneficial for future alloy design to address this problem. In this work, synchrotron X-ray diffraction during stress relaxation experiments was used to characterise the time dependent plastic behaviour of two commercially pure titanium samples (grade 1 and grade 4) with different oxygen content at 4 different temperatures (room temperature, 75 $^{\circ}$C, 145 $^{\circ}$C and 250 $^{\circ}$C). Lattice strains were measured by tracking the diffraction peak shift from multiple crystallographic plane families (21 diffraction rings) as a function of their orientation with respect to the loading direction. Critical resolved shear stress, activation energy and activation volume were established for both prismatic and basal slip as a function of temperature and oxygen content by fitting a crystal plasticity finite element model to the lattice strain relaxation responses measured along the loading axis for five strong reflections. Higher strain rate sensitivity was found to lead to higher plasticity during cold dwell.
\end{abstract}


\end{frontmatter}

\section{Introduction}
\label{Intro}
Titanium alloys are widely used for components within the cooler front section of aeroengines such as fan blades and compressor discs due to their low density and high strength at low temperature~\cite{CUDDIHY2017,LITTLEWOOD2012}. Although the operational temperature of the Ti alloy parts in such application is low (typically below 200~$^{\circ}$C), they can be rate-sensitive under strain and stress-controlled loading (stress relaxation and creep), an effect which is strong in some (but not all) Ti alloys~\cite{ZHANG2016,NEERAJ2000}. This phenomenon must be considered during in-service conditions in aero engines as there is an extended stress dwell during the cruise phase of a flight cycle. The successive accumulation of plastic damage is commonly referred to as ‘cold dwell fatigue’ and results in the drastic reduction in the lifetime of Ti and Ti-alloy components~\cite{BACHE2003,DUNNE2008}. This problem has been studied for several decades, however, there still exist several aspects of the micromechanical behaviour that remain unknown.

Cold dwell fatigue is mainly manifested in the $\alpha$ phase (HCP) Ti~\cite{CONRAD1981,CONRAD2011,PATON1976Deformation,WILLIAMS2002}, and is highly sensitive to the crystallographic orientation of grains. Post-mortem fatigue crack analysis indicates  that dwell fatigue is enhanced by loading along the \{0002\} crystal plane normal direction, where the crack surface was found to be associated with the formation of near the \{0002\} facet and lies nearly perpendicular to the loading direction~\cite{SINHA2004,SINHA2006_2,SINHA2006}. This is supported by a classical Stroh model for crack nucleation~\cite{STROH1954} and a qualitative model introduced by Evans \& Bache~\cite{EVANS1994,BACHE2003_2}. A more recent mechanistic study by Dunne \emph{et al}~\cite{DUNNE20071061,DUNNE2007Proceedings} suggested that cold dwell effect is controlled by load shedding between ‘soft’ and ‘hard’ grain pairs, where the ‘soft’ grain has its crystallographic \textit{c}-axis approximately perpendicular to the loading direction (well orientated for slip) while a ‘hard’ grain has its \textit{c}-axis approximately parallel to the loading direction (badly orientated for slip). During a stress dwell, localised time dependent plasticity of easily activated slip systems (basal and prismatic slip) in ‘soft’ grains leads to load shedding which in turn causes a large increase in stress within the neighbouring ‘hard’ grain, where the facets  are observed to initiate. These hard-soft grain neighbour configurations are termed ‘rouge grain pairs’. The magnitude of stress that builds along their shared grain boundary during a stress dwell is known to vary with the ‘soft’ grain crystal orientation~\cite{DUNNE2008}. To have a thorough understanding of dwell fatigue, it is necessary to understand the different deformation behaviour for grains with different crystal orientations (i.e. ‘hard’ and ‘soft’ grain orientations) and especially the plastic deformation during the stress dwell period. To achieve this, it is important to quantify the plasticity processes as a function of time. Specifically, the dislocation processes on prismatic and basal slip systems with respect to their critical resolved shear stresses must be described.

The majority of studies that have investigated cold dwell fatigue have been performed at ambient temperature. However, there is clear evidence showing that the cold dwell effect is also sensitive to temperature, noting that the effect is not observed at high temperature~\cite{TITANIUM}. Zhang \emph{et al}~\cite{ZHANG2015} utilised a crystal plasticity model to show that the worst-case scenario temperature in Ti6Al alloy is around 120 $^{\circ}$C and is diminished when the temperature exceeds 230~$^{\circ}$C. To understand this effect in greater detail, the effect of temperature on dislocation activity of different slip systems needs to be understood and quantified in the context of cold dwell fatigue. In addition, oxygen is a common interstitial in titanium alloys. Typically, in high strength grade Ti alloys, oxygen is intentionally added in order to improve the strength, corrosion and wear resistance of the alloy~\cite{STRINGER1960,KALE2019}. However, in most of Ti alloys, oxygen is unintentionally introduced during its raw metal processing or due to service conditions~\cite{Yu2015}. Although oxygen is present in almost all Ti alloys, the role of oxygen on time dependent plasticity and cold dwell fatigue is not clear; a better understanding is highly attractive as it may guide future alloy design strategies and impurity control to suppress the cold dwell effect.  

Strain rate sensitivity (SRS) is a crucial factor that governs the load shedding phenomenon~\cite{JUN2016NANO}. As titanium alloys are highly anisotropic at the grain scale, both elastically and plastically~\cite{BRITTON2010}, SRS is likely to vary depending on the grain orientation, which is due to the difference in SRS for different slip systems~\cite{BRITTON2015}. This was confirmed by Jun \emph{et al}~\cite{JUN2016NANO} using nanoindentation, showing grain orientation dependent rate sensitivity in Ti6242. Jun \emph{et al}~\cite{JUN2016} continued this study using in-situ compression testing of micro-pillars manufactured from the $\alpha$ phase of Ti6242, which successfully isolated the behaviour of each slip system. Significant slip system dependent rate sensitivity was observed where prism slip has a significantly higher strain rate sensitivity exponent, \textit{m}, over basal slip. Although abundant research on strain rate sensitivity have been performed and the hypothesis has been verified, interpreting the \textit{m} value and correlating it to physical phenomena is still an open question~\cite{JUN2016}. 

More recently, there have been several studies on the deformation of $\alpha$ titanium using synchrotron diffraction~\cite{STAPLETON2008,WARWICK2012,CHATTERJEE2017,WIELEWSKI2017,DAWSON2018,SOFINOWSKI2019,ZHANG2020}. Such experiments enable direct assessment of lattice strains from grain families with common grain orientations from a polycrystal sample. By measuring the lattice strain response, it is possible to infer the behaviour of different slip systems.  

In the current study, time dependent plasticity (stress relaxation) behaviour at different temperatures up to 250~$^{\circ}$C of two commercially pure titanium (CP-Ti) with different oxygen content was observed by synchrotron X-ray diffraction. Measured lattice strains from multiple lattice plane families were compared with and calibrated to simulated lattice strains from crystal plasticity finite element (CPFE) simulations using methods established in prior work~\cite{ERINOSHO2016}. The objective of this work was to quantify the key parameters (activation volume, activation energy and critical resolved shear stress) controlling the time dependent plasticity of two grades of CP-Ti alloys to reveal the effect of temperature and oxygen on the cold dwell fatigue phenomenon. The strain rate sensitivity exponent, \textit{m}, is often used to describe rate effects though \textit{m} itself can vary significantly with strain rate. Here, we determine values of \textit{m} (over a defined strain rate range) for individual lattice plane families and for an individual slip system, which aims to correlate the m value with physical phenomena, in context of cold dwell fatigue.  

\section{Materials and method}
\subsection{Materials}
Two types of commercially pure titanium (grade 1 and grade 4) with different oxygen content were selected in this experiment, where the grade 1 material was supplied in the form of a rolled sheet while the grade 4 material was supplied in the form of a rolled bar. The composition of these two materials are shown in Table~.\ref{table.1}. 
Microstructural analyses were conducted on these samples. The specimens were metallographically prepared with abrasive media, using SiC papers (up to 4000 grit), and a final polish with $\approx$ 50~nm colloidal silica. Electron backscatter diffraction (EBSD) maps were obtained using a Zeiss Merlin scanning electron microscope (SEM) equipped with a Bruker e-flash detector, operating at an accelerating voltage of 20~kV and a probe current of 20~nA. The microstructure of the materials are shown in Fig.~\ref{fig.1}(a) and Fig.~\ref{fig.1}(b), with both grade 1 and grade 4 having equiaxed grains with an average grain size of 30~$\mu$m and 17~$\mu$m respectively.
\begin{figure}[b!]
\centering
\subfigure{
\includegraphics[width=.95\textwidth]{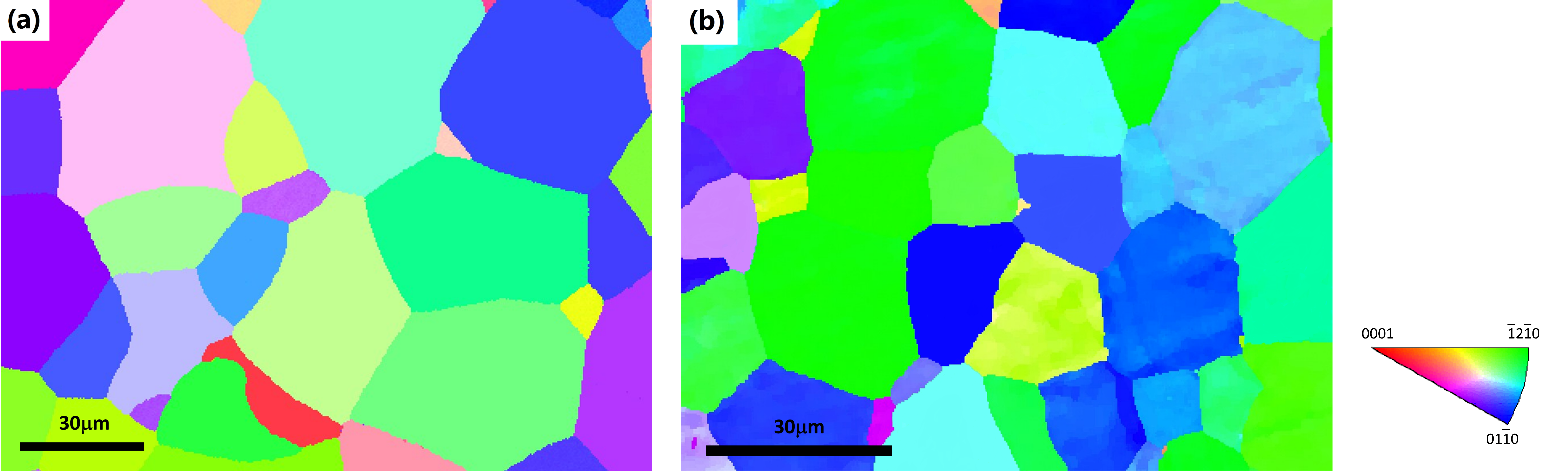}
}
\\
\subfigure{
\includegraphics[width=.9\textwidth]{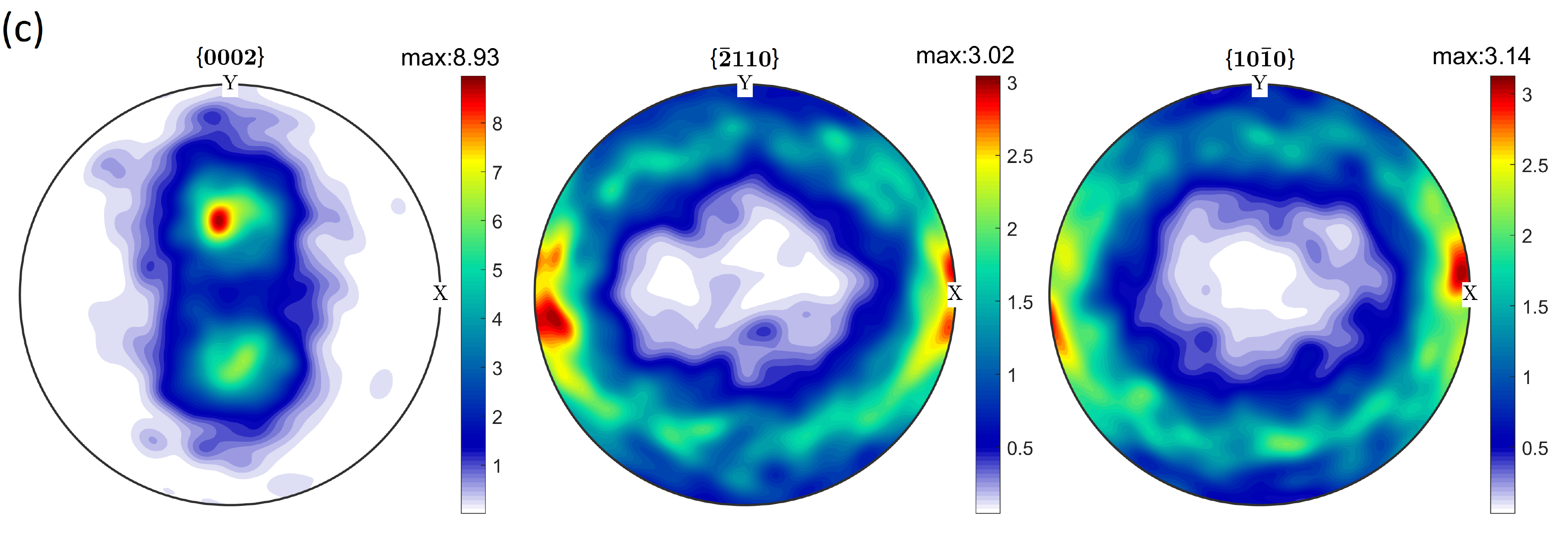}
}\\
\subfigure{
\includegraphics[width=.9\textwidth]{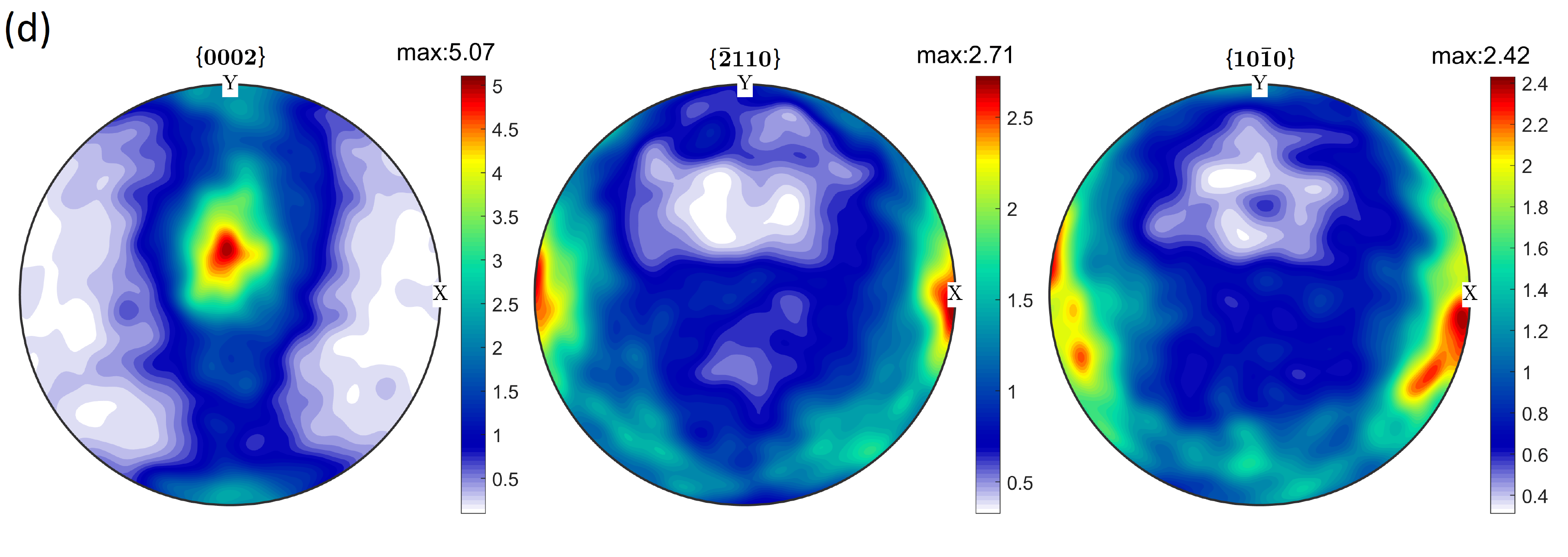}
}
\phantomcaption
\end{figure}
\begin{figure}[t!]\ContinuedFloat
\centering
\subfigure{
\includegraphics[width=.9\textwidth]{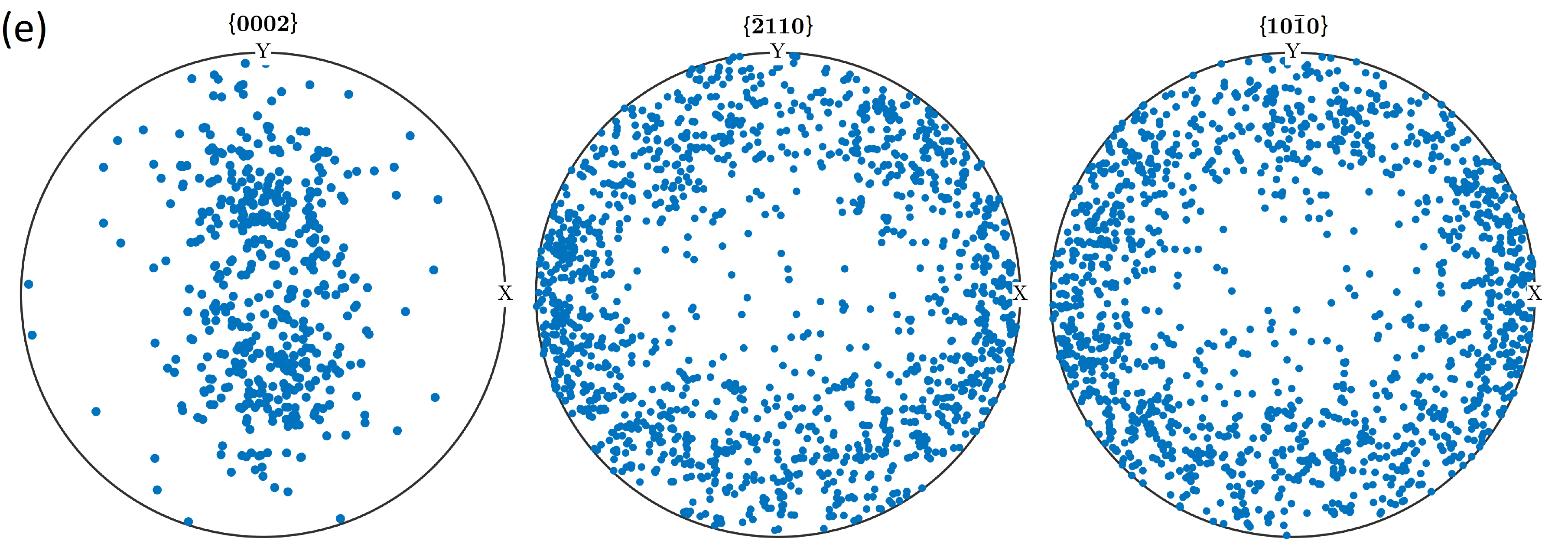}
}\\
\subfigure{
\includegraphics[width=.9\textwidth]{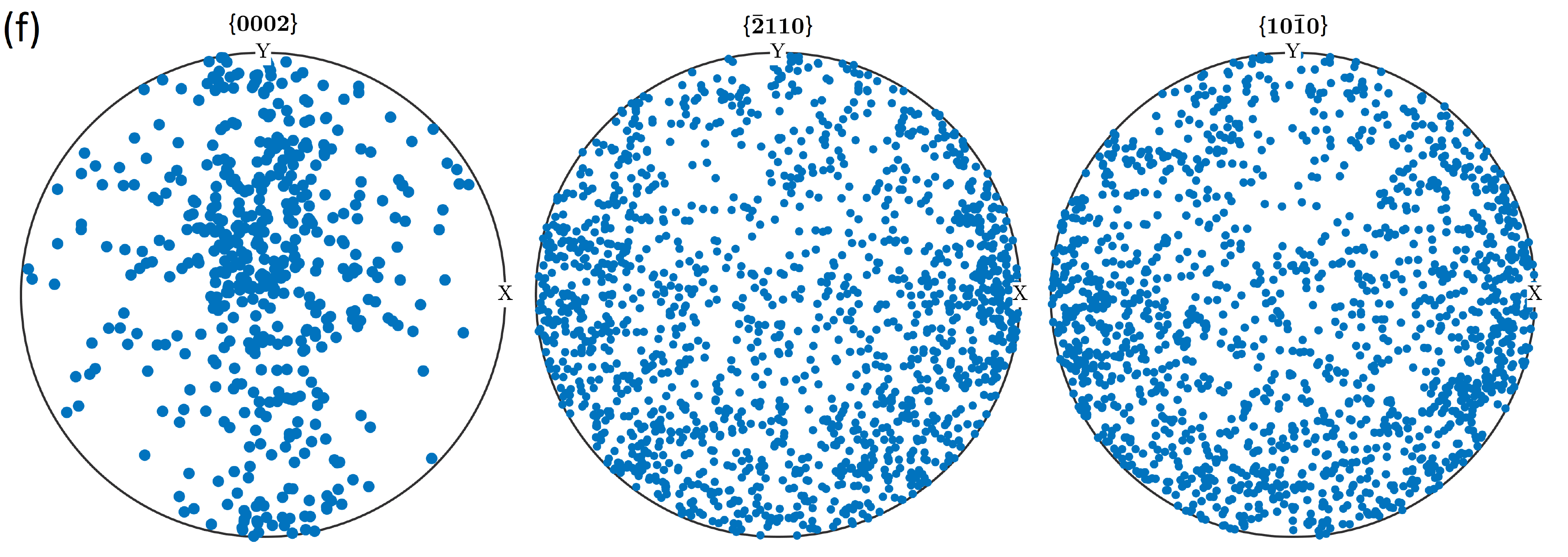}
}
\caption{(a) EBSD map of CP-Ti grade 1 (IPF colour map along the rolling direction of the raw material); (b) EBSD map of CP-Ti grade 4 (IPF colour map along the rolling direction of the raw material) showing the grain morphology and grain size; (c) Pole figures of the CP-Ti grade 1 (X along the rolling direction, which was the loading direction); (d) Pole figures of the CP-Ti grade 4 (X along the rolling direction, which is the loading direction)  with 10 degree of width used for contours calculation; (e) Scatter pole figure plots of the polycrystal finite element model for CP-Ti grade 1; (f) Scatter pole figure plots of the polycrystal finite element model for CP-Ti grade 4.}
\label{fig.1}
\end{figure}
The texture of the materials are shown in the pole figures in Fig.~\ref{fig.1}(c) and Fig.~\ref{fig.1}(d), which were obtained from a larger area containing over 3500 grains. There is a preference for the $c$-axis to be nearly perpendicular to the rolling direction, which is common in hot-rolled titanium alloys~\cite{Larson1974}. In order to ensure the activation of the easy slip system (prismatic) during stress relaxation experiments, Electro-Thermal-Mechanical Tester (ETMT) samples were cut with the loading axis along the rolling direction of the rod and the sheet, the X direction in these pole figures. With the predominant basal \{0002\} plane aligned approximately perpendicular to the tensile axis (X in the pole figures) in these specimens, the calculated Schmid factor for basal slip was calculated to be low, encouraging prismatic slip as the dominant slip system in the experiment. The tensile sample was 52~mm in length, 1~mm in thickness with a 16~mm gauge length and a 2~mm gauge width. 
\FloatBarrier
\begin{center}
\begin{table*}[ht]
 \begin{tabular*}{1\textwidth}{ @{\extracolsep{\fill}}c@{\extracolsep{\fill}}c@{\extracolsep{\fill}}c@{\extracolsep{\fill}}c@{\extracolsep{\fill}}c@{\extracolsep{\fill}}c@{\extracolsep{\fill}}} 
 \hline
 & Ti & Fe~(wt.~\%) & O~(wt.~\%) & N~(wt.~\%) & Others total~(wt.~\%)\\
  \hline
 Grade 1 & Balance & $<$0.05 & 0.08 & 0.02 & 0.2 \\ 
 \hline
 Grade 4 & Balance & 0.05-0.055 & 0.32 & 0.006 & 0.4 \\ 
 \hline
\end{tabular*}
\caption{Composition of the two CP-Ti raw materials.}
\label{table.1}
\end{table*}
\end{center}
\FloatBarrier
\subsection{Experiment}
\begin{figure}[htb!]
\centering
\includegraphics[width=1\textwidth]{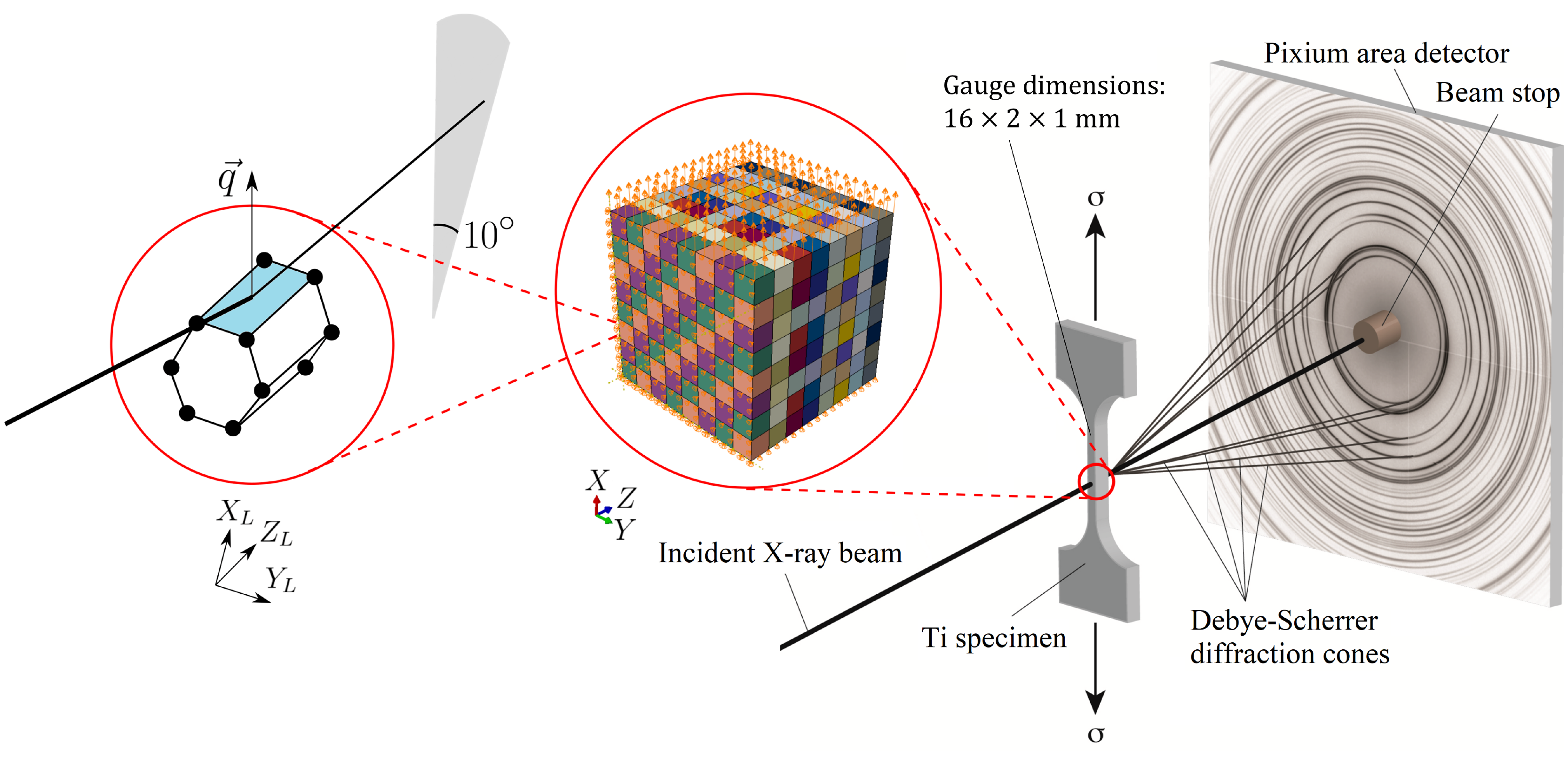}
\caption{Sketch of the synchrotron diffraction experiment setup, the loading direction is along the rolling direction of the two CP-Ti alloys. Finite element mesh of the RVE used with the CPFE model, with arrows indicate the uniaxial boundary conditions and an example of a grain with its prismatic plane satisfied the diffraction condition in vertical (load axis) sector.}
\label{fig.2}
\end{figure}

A set of stress relaxation experiments were carried out on beamline I12~\cite{Drakopoulos2015} at the Diamond Light Source. An illustration of the experiment setup is shown in Fig.~\ref{fig.2}. 
The beamline was configured for diffraction, operating with a monochromatic beam at 79.79~keV, calibrated with a $\textrm{CeO}_2$ standard, and an incident beam size of 1 x 1~$\textrm{mm}^2$. Dogbone shaped Ti specimens were placed in the path of the incident beam, enabling Debye Scherrer diffraction rings to be collected in transmission, imaged with a 2D Pixium area detector. The exposure time for each frame was 1 second and therefore the acquisition rate for the diffraction rings was 1 frame per second. The sample to detector distance was set and measured to be 1097~mm from the standard sample. 

The samples were deformed and heated on an Instron ETMT. Here, samples were heated through a DC current that were fixed between water cooling grips, that gives a parabolic temperature distribution~\cite{Roebuck2001,Roebuck2004,SULZER2018}. Temperature measurement and control was achieved from an R-type thermocouple spot welded to the centre of each sample gauge. To ensure the temperature of the sample volume probed by the X-rays was known, the samples were centred on the incident beam path and positioned using radiographic imaging such that the sample volume was in close proximity to the thermocouples. Tests were carried out at four different temperatures: room temperature, 75~$^{\circ}$C, 145~$^{\circ}$C and 250~$^{\circ}$C. The 250~$^{\circ}$C test of CP-Ti grade 1 was abandoned due to excessive sample softening at this temperature.  At each designated temperature, samples were heated and held idle for 30 seconds to allow for temperature stabilisation, followed by deformation whilst collecting Debye Scherrer diffraction data. Each diffraction pattern was recorded as a 16-bit image and synchronised with the mechanical data of the ETMT for offline analyses. 

The Ti samples were subjected to pre-determined load-hold cycles. Prior to in-situ testing, target loads were obtained from trial tests. The target loads corresponded to a stroke displacement at which the yield stress had just been exceeded. For in-situ testing, the samples were loaded at a constant load rate until the macroscopic yield point had been exceeded by a small plastic strain, achieved in the range of 120~s and 150~s. To within slight differences between samples, these corresponded to macroscopic strain rates of approximately $3\times10^{-5}~\textrm{s}^{-1}$ (as measured from the ETMT stroke rate) within the initial elastic part of the loading phase. Once the target load had been reached, the sample displacement (and thus the total strain) was then held constant for a period of 5 minutes, during which the macroscopic stress was found to relax. After a 5 minutes strain-hold, the load was incremented to a level that was slightly below the peak load used at the start of the previous stress relaxation cycle. This load increment procedure follows that of Wang~\emph{et al}~\cite{WANG2006} and corresponds to an elastic deformation of the sample that permits comparison of stress relaxation at two stress levels, but critically, any change in dislocation substructure is negligible. The sample was again held at a constant displacement for a further period of stress relaxation. This load and hold cycle was repeated for a total of 5 stress relaxation periods with diffraction patterns collected throughout (shown in Fig.~\ref{fig.6}(a) and~\ref{fig.6}(b)). The four elastic reload were introduced to verify that change in dislocation density, $\rho$,  remains relatively small during these relaxation periods. 

\begin{figure}[hbt!]
    \centering
    \includegraphics[width=0.75\textwidth]{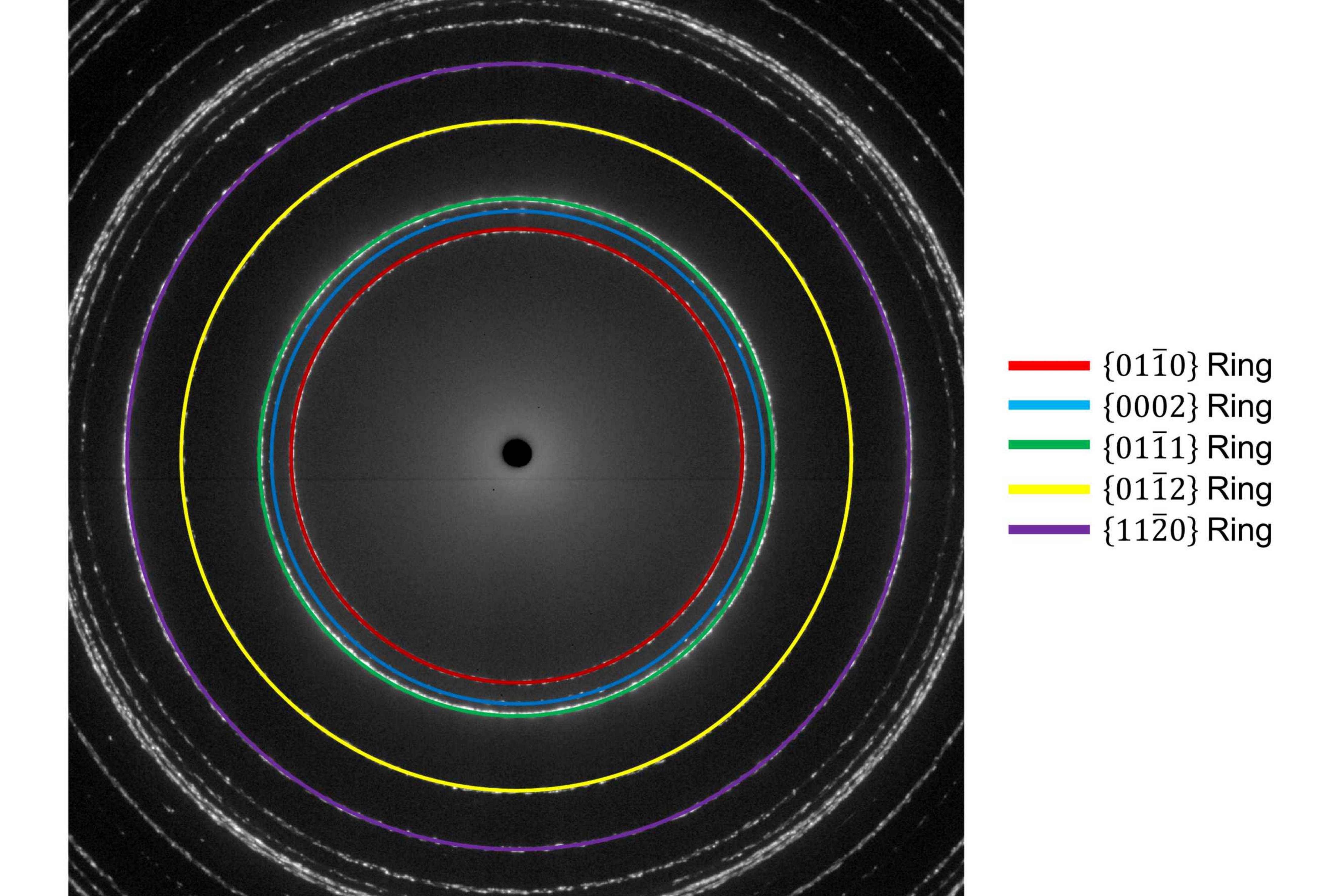}
    \caption{Debye Scherrer diffraction patterns of CP-Ti grade 1 at room temperature (the vertical direction of the diffraction patterns is the rolling direction and the horizontal direction is the transverse direction of the sheet CP-Ti grade 1), first 21 rings with ascending ring radius were fitted with ellipses (first 5 rings were used here as examples).}
    \label{fig.3}
\end{figure}

Diffraction pattern data reduction was achieved using Data Analysis WorkbeNch (DAWN)~\cite{BASHAM2015}. 
As shown in Fig.~\ref{fig.3}, the 2D diffraction rings were fitted with ellipses, with minor radii aligned near parallel to the vertical axis (tensile direction) of the diffraction pattern, and major radii was aligned near parallel to the horizonal axis (transverse direction) of the diffraction pattern. 
The radius and thus the scattering vector, \textit{q}, at any azimuthal angle around a diffraction ring could then be found from the size and alignment of the major and minor axes. Ellipse fitting was applied to the first 21 rings with ascending ring radius. For further details of the data reduction process, the reader is referred to Filik~\emph{et al}~\cite{Filik2017}. 

\begin{figure}[hbt!]
\centering
\subfigure{
\includegraphics[width=.435\textwidth]{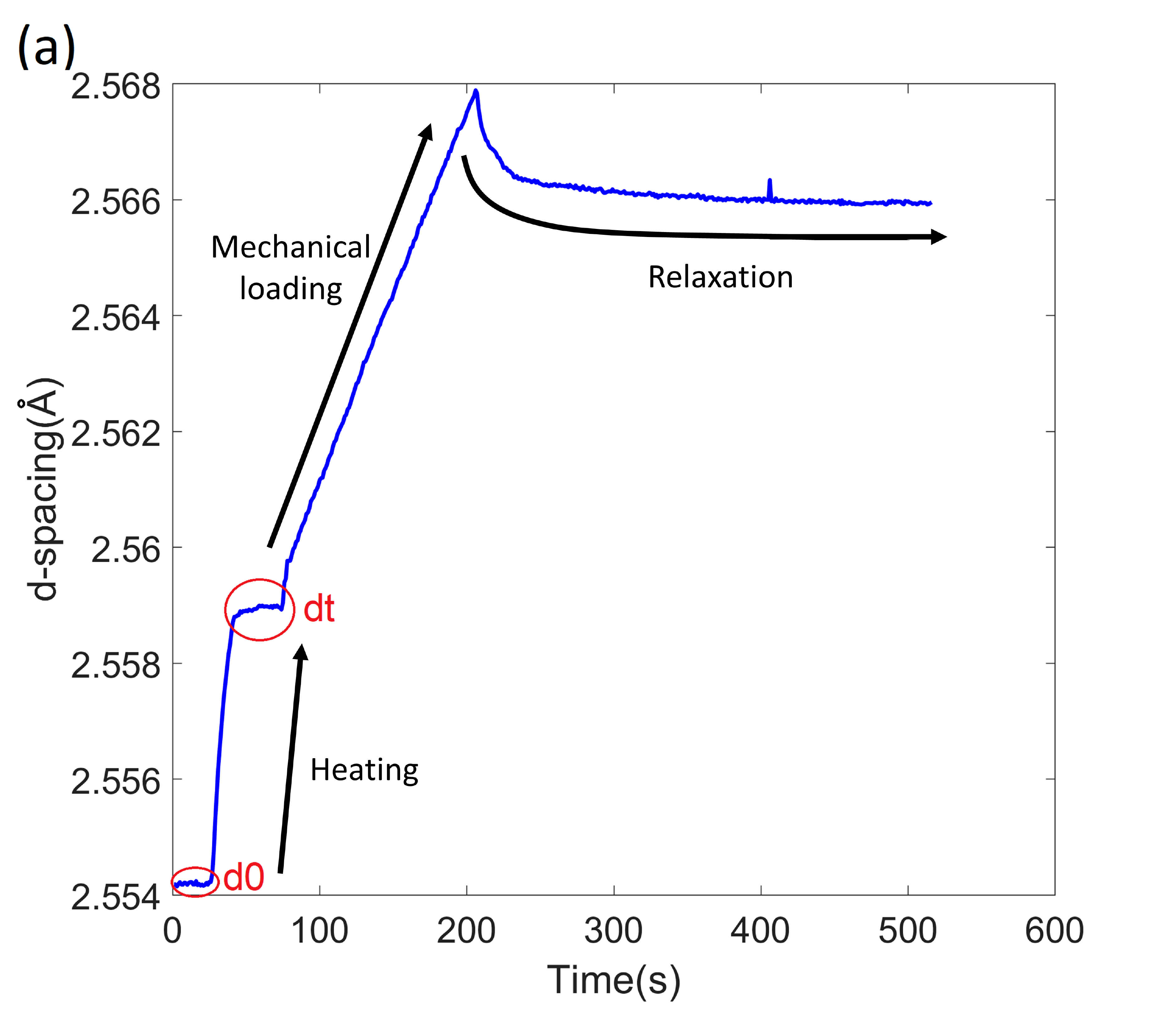}
}
\subfigure{
\includegraphics[width=.45\textwidth]{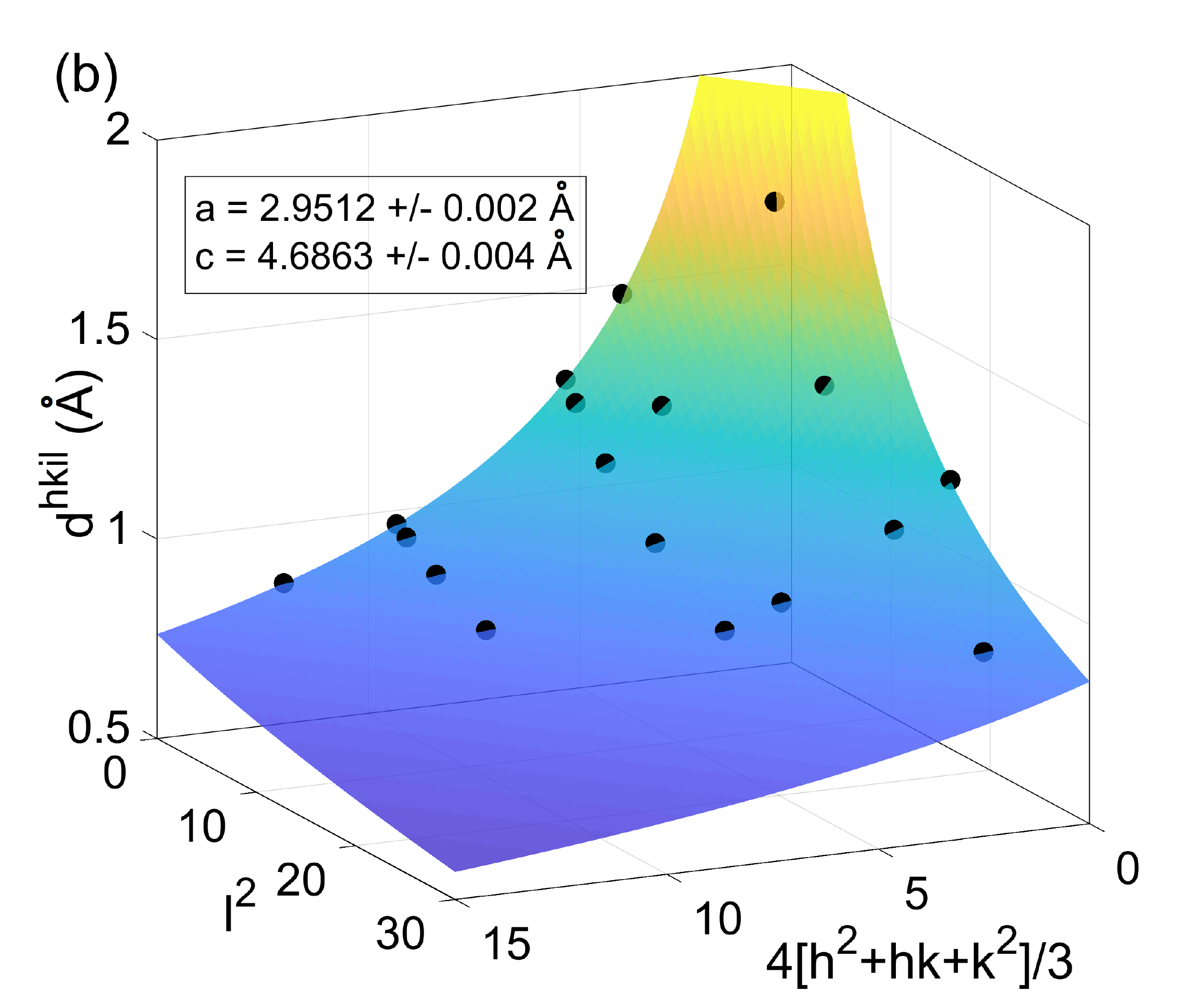}
}\\
\caption{(a) Determination of $d_0$ and $d_t$ using the two periods of idle for each plane family ($\{01\bar{1}0\}$ plane in CP-Ti grade 4 at 145~$^{\circ}$C as example); (b) Non-linear regression fit the crystal geometry equation of HCP (CP-Ti grade 4 sample at room temperature as example).}
\label{fig.4}
\end{figure}

Time-dependant lattice spacings, $d^{hkil}$, were obtained through: $d^{hkil}=2\pi/q$. Lattice strain, $\epsilon^{hkil}$, keeps changing as sample was deformed and was subsequently calculated as: $\epsilon^{hkil}=(d^{hkil}-d_0^{hkil})/d_0^{hkil}$ for each second, where $d_0^{hkil}$ is the initial lattice spacing for plane \textit{hkil}. Fig.~\ref{fig.4}(a) shows the lattice strain evolution during heating, loading and first relaxation cycle of CP-Ti grade 4 at 145~$^{\circ}$C.
It is found that the lattice d-spacing only increases by less than 0.01~\si{\angstrom} relative to zero load during the whole test, highlighting the importance in determining the initial lattice spacing $d_0^{hkil}$ in order to calculate lattice strain. To do this, a consistent set of stress free lattice parameters $a$ and $c$ needs to be determined, and for each diffraction ring, a $d_0$ value was obtained by averaging the d-spacing values during the first idle period at zero load and room temperature (as shown in Fig.~\ref{fig.4}(a)). These $d_0$ values from each diffraction ring (hence each diffraction plane) were then averaged for both vertical and horizontal directions (assuming at zero load the diffraction ring is perfect circle), which yield $d^{hkil}$ at zero load. These $d^{hkil}$ values for 21 different plane families were fitted into the crystal geometry equation for HCP:
\begin{equation}\label{eq1}
    \frac{1}{d^{hkil}} =\frac{3(h^2+hk+k^2)}{4a^2}+\frac{l^2}{c^2}  
\end{equation}
where $a$ and $c$ are the lattice parameters of HCP crystal, $h$, $k$ and $l$ are the Miller Indices of a crystal plane. Therefore, the non-deformed lattice parameters, $a_0$  and $c_0$ of the CP-Ti samples can be found (see Fig.~\ref{fig.4}(b)). The initial stress free lattice spacing, $d_0^{hkil}$, for each plane family was then obtained by substituting values of $a_0$ and $c_0$ into the formula. 

\subsection{Crystal plasticity model}
The CP-Ti materials were modelled as a single phase, $\alpha$, with a HCP crystal structure. The elastic parameters at four different temperatures are indicated in Table.~\ref{table.2}. The thermal expansion coefficients used in the crystal plasticity model are stated in Table.~\ref{table.3}. All of these parameters were taken from literature values. 
\FloatBarrier
\begin{center}
\begin{table}[h]
 \begin{tabular*}{1\textwidth}{ c@{\extracolsep{\fill}}c@{\extracolsep{\fill}}c@{\extracolsep{\fill}}c@{\extracolsep{\fill}}c@{\extracolsep{\fill}}c@{\extracolsep{\fill}}c} 
 \hline
 %\multicolumn{7}{l}{Moduli~(GPa) and Poisson’s ratio}\\
 %\hline
 & $E_{11}$(GPa)  & $E_{33}$(GPa) & $G_{12}$(GPa) & $G_{13}$(GPa) & $\nu_{12}$ & $\nu_{13}$ \\ \hline
 RT & 84.7 & 118.4 & 28.6 &	39.5 & 0.46 & 0.22\\ 
 75~$^{\circ}$C & 80.5 & 113.8 & 27.5 &	37.9 & 0.47 & 0.22\\
 145~$^{\circ}$C &	75.1 & 106.1 & 25.5 & 35.2 & 0.47 & 0.23\\
 250~$^{\circ}$C & 66.9 & 94.6 & 22.6 &	31.2 & 0.48 & 0.23\\
 \hline
\end{tabular*}
\caption{Parameters for polycrystalline Ti crystal plasticity model under uniaxial loading at 4 different temperatures~\cite{DUNNE20071061,ZHANG2015}}
\label{table.2}
\end{table}
\end{center}
\FloatBarrier
\FloatBarrier
\begin{center}
\begin{table}[h]
\begin{tabular*}{0.825\textwidth}{ c@{\extracolsep{\fill}}c@{\extracolsep{\fill}}c@{\extracolsep{\fill}}c} 
\hline
Coefficient  & 11 & 22 & 33 \\ \hline
$\alpha(K^{-1})$ & $1.8\times10^{-5}$ & $1.8\times10^{-5}$ & $1.1\times10^{-5}$\\ 
\hline
\end{tabular*}
\caption{Single crystal thermal expansion tensor coefficients for Ti~\cite{CONRAD2011,PATON1976Deformation}.}
\label{table.3}
\end{table}
\end{center}
\FloatBarrier
Dislocation slip occurs predominantly with $\overrightarrow{a}$ type Burgers vector on the basal and prismatic slip planes in Ti alloys~\cite{NEERAJ2000,WILLIAMS2002,SURI1997}. This is due to the critical resolved shear stress (CRSS) magnitude required to activate $\overrightarrow{a}$ or $\overrightarrow{c}+\overrightarrow{a}$ type Burgers vector dislocations on a $1^{st}$ or $2^{nd}$ order pyramidal plane being approximately three time higher than the CRSS for basal slip at room temperature~\cite{DUNNE2008,PATON1976Deformation}. As the macroscopic yield stress of each sample was only just exceeded, only limited pyramidal type slip was expected due to its difficult activation.
In order to simulate the lattice strain evolution during the stress relaxation cycles, a crystal plasticity finite element model was built~\cite{DAS2018}. The deformation gradient is decomposed multiplicatively into elastic and plastic parts:
\begin{equation}\label{eq2}
    \bm{F} = \bm{F_{e}F_{p}}
\end{equation}

The shear strain rate on each individual slip system is determined through a constitutive law which is physically developed by Dunne \emph{et al}~\cite{DUNNE20071061}, in which the slip rate, $\dot{\gamma}^{\kappa}$, on a slip system $\kappa$ is given by:
\begin{equation}\label{eq3}
    \dot{\gamma}^{\kappa} = {\rho}{b^{\kappa}}^{2}{\nu}\exp\left(-\frac{\Delta{F}^{\kappa}}{k_BT}\right)\sinh{\left(\frac{|{\tau^{\kappa}}-{\tau_{c}^{\kappa}}|{\Delta{V}}^{\kappa}}{k_BT}\right)}\text{sgn}(\tau^{\kappa})
\end{equation}
where: $\rho$ is the density of gliding dislocations; $\nu$ is the jump frequency (i.e. attempts of dislocations to jump energy barriers); $k_B$ is the Boltzmann constant. All parameters above remain constant in the slip law and the magnitudes are summarised in Table.~\ref{table.4}. $T$ is the absolute temperature and $b^{\kappa}$ is the magnitude of the Burgers vector of the slip system $\kappa$ (calculated from the $a$ and $c$ values measured above).

The remaining three terms are material specific parameters that govern dislocation motion and are the targeted unknowns we seek to determine: $\Delta{F}^{\kappa}$ is the thermal activation energy for $\kappa$ slip system, $\Delta{V}^{\kappa}$ is the activation volume, which refers to the volume that a dislocation is swept when it passes obstacles~\cite{DUNNE20071061} ($\Delta{V}^{\kappa}$ is typically in the range 1-100$b^3$), and $\tau_c^{\kappa}$ is the critical resolved shear stress (CRSS) for slip system $\kappa$. Summing up these shear strain rates over the possible crystallographic slip systems allows the plastic velocity gradient to be determined~\cite{DUNNE20071061}:
\begin{equation}\label{eq4}
    \bm{L_p} = \sum\nolimits_{\kappa} \dot{\gamma}^{\kappa} \bm{s^{\kappa}} \otimes \bm{n^{\kappa}}
\end{equation}
where $\bm{s^{\kappa}}$ is the unit slip direction and $\bm{n^{\kappa}}$ is the unit normal to the slip plane of the $\kappa^{th}$ slip system.
\FloatBarrier
\begin{center}
\begin{table}[h]
\begin{tabular*}{0.4\textwidth}{c@{\extracolsep{\fill}}c}
\hline
$\rho$ & 5~$\mu\text{m}^{-2}$ \\
$\nu$ & $10^{11}$~Hz  \\
$k_B$ &  $1.38 \times 10^{-23}~\text{JK}^{-1}$\\
\hline
\end{tabular*}
\caption{Values of the fixed parameters in the slip law~\cite{DUNNE20071061,ZHENG2016}}
\label{table.4}
\end{table}
\end{center}
\FloatBarrier
Polycrystal simulations with 512 ($8~\times~8~\times~8$) cubic grains were carried out using the commercial finite element (FE) software Abaqus 2016. Each grain consists of 64 ($4~\times~4~\times~4$) quadratic elements (C3D20R). The sample grain orientations (Euler angles), as measured by EBSD (Fig.~\ref{fig.1}(b)  and Fig.~\ref{fig.1}(c)), were converted into rotation matrices, $\bm{R}$, then selected randomly and assigned to each orientation grain by grain in the model. Therefore, the overall texture of the model was approximately equal to that of the real sample, so that the effect of texture on lattice strains was minimised~\cite{XIONG2020}. The boundary conditions are illustrated in Fig.~\ref{fig.2}. Three stages were implemented to simulate the stress relaxation experiment. In all three stages, the X=0, Y=0 and Z=0 surfaces were constrained in the X, Y and Z directions respectively. During the heating stage, temperatures were raised to each target temperature in 30~s whilst the other three surfaces were free to expand. During the loading stage, a uniaxial displacement was applied to the X=1 surface over the same period of time as the experiments (with a corresponding strain rate of \textit{ca}.~$3~\times~10^{-5}$ $\text{s}^{-1}$). During the relaxation stage, the displacement of the right surface was held fixed along the load axis (X axis) for 300~s. 

In order to compare the simulation result with the lattice strain measured by diffraction, the Green-Lagrange elastic strain tensor, $\bm{E_e}$, is rotated to the lattice orientation frame (${X_L}$, $Y_L$ and $Z_L$) through the rotation matrix $\bm{R}$ to give the lattice strain $\bm{E_e^0}$  as follows~\cite{GRILLI2020}:
\begin{equation}\label{eq5}
    \bm{E_e^0} = \bm{R^TE_eR} = \frac{1}{2}\bm{R^T(F_e^TF_e-I)R}
\end{equation}
As illustrated in Fig.~\ref{fig.2}, the \{01$\bar{1}$0\} planes diffract along the vertical axis with diffraction vectors, $q$, approximately parallel (with $10^{\circ}$ tolerance) to the load axis. For grains that satisfy this condition, strain along the scattering vector, $q$, was extracted from the elastic strain $\bm{E_e^0}$ matrix and then averaged from all the elements within these grains and directly compared with the experimental lattice strains. The five crystal plane families with the highest diffraction intensities along the loading direction were selected for comparison, their high intensity implying that a large number of grains satisfied the diffraction condition, ensuring more reliable results. These five crystal plane families were the prismatic plane \{01$\bar{1}$0\}, first ordered pyramidal plane \{01$\bar{1}$1\}, second ordered pyramidal planes; \{11$\bar{2}$2\}, \{02$\bar{2}$1\} and \{12$\bar{3}$1\} for grade 1. For grade 4, these were the prismatic plane \{01$\bar{1}$0\}, first ordered pyramidal plane \{01$\bar{1}$1\}, second ordered pyramidal planes, \{11$\bar{2}$2\}, \{02$\bar{2}$1\} and \{01$\bar{1}$2\}. Another important crystal plane \{0002\}, the basal plane, was not deemed to be available in either grade 1 or grade 4 due to the sample texture and the experimental setup with the incident beam along ND (recall Fig.~\ref{fig.1}(a) and~\ref{fig.1}(b)). 
\section{Results and discussion}
\subsection{Thermal expansion and elastic deformation}
\begin{figure}[hbt!]
\centering
\includegraphics[width=1\textwidth]{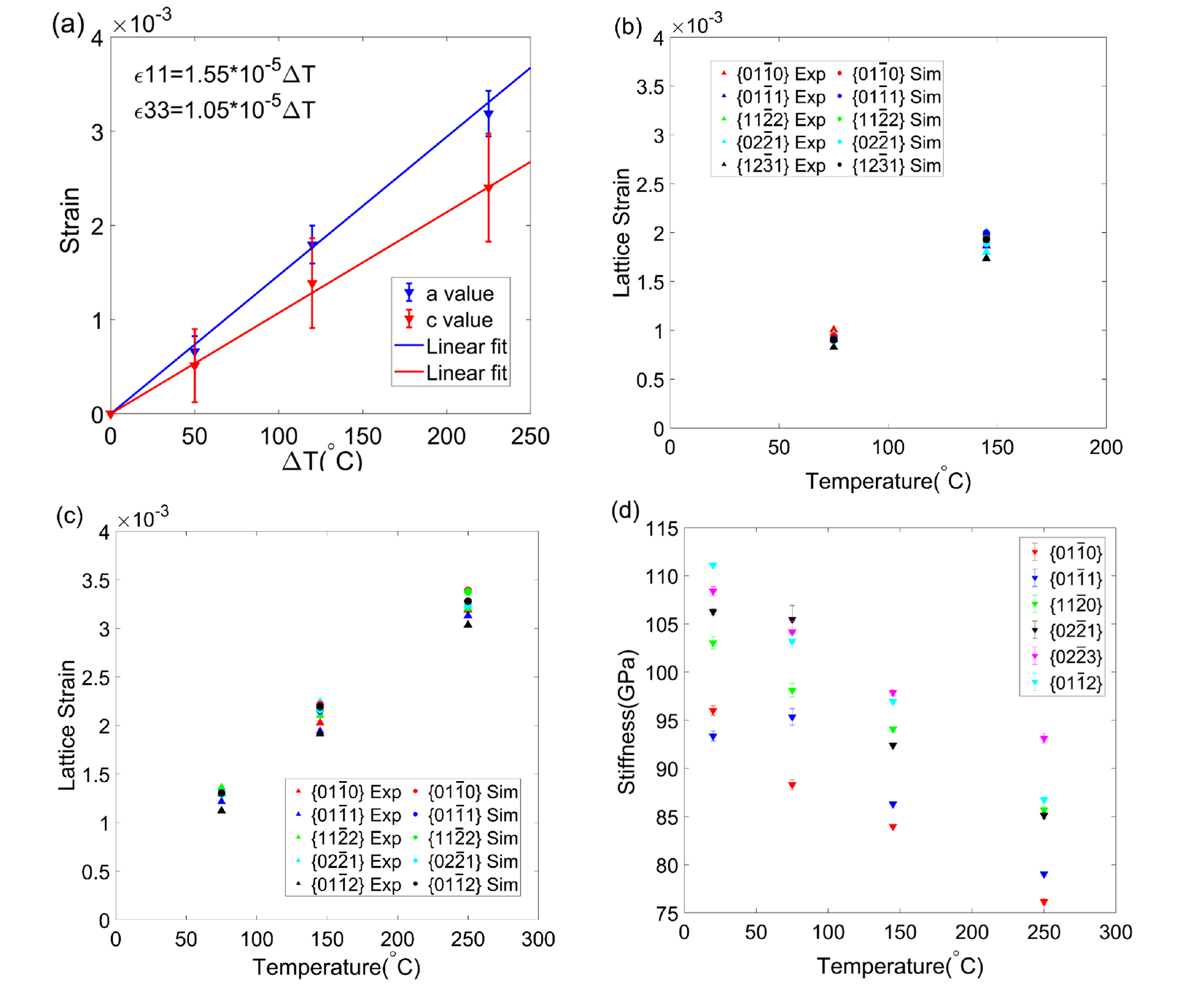}
\caption{(a) Thermal expansion strain in the $a$ and $c$ directions, the gradient of the linear fitted lines enables a measurement of the thermal expansion coefficient (CP-Ti grade 4 samples); (b) Comparison of the thermal expansion strain of the experiment and simulation of CP-Ti grade 1 along the loading direction; (c) Comparison of thermal expansion strain of the experiment and the simulation of CP-Ti grade 4 along the loading direction; (d) Variation of the stiffness of 6 plane families vs. temperature in CP-Ti grade 4. }
\label{fig.5}
\end{figure}

The lattice parameters at elevated temperatures, $a_t$  and $c_t$, can be calculated using the method for calculating $a_0$  and $c_0$  values as discussed in section 2.2, but instead using the d-spacing values in the second idle period. A series of $a_t$ and $c_t$ values are therefore determined at each target temperature. The increase in lattice spacing between $a_0$, $c_0$ and $a_t$, $c_t$ are purely from thermal expansion as the ETMT load was set to zero in a feedback loop, therefore no force is applied to the samples. Within each sample, the lattice strain caused by thermal expansion can then be calculated as follows:  
\begin{equation}\label{eq6}
    \epsilon_{11} = (a_t-a_0)/a_0
\end{equation}
\begin{equation}\label{eq7}
    \epsilon_{33} = (c_t-c_0)/c_0
\end{equation}
Fig.~\ref{fig.5}(a) shows the strain in lattice parameters against changing temperatures of CP-Ti grade 4 samples, the gradients of the linear fitted lines represent the thermal expansion coefficient in the $a$ and $c$ directions of the Ti HCP crystal structure. 
It was found that the thermal expansion coefficients in the $a$ direction is higher than that of the $c$ direction, which agrees with values reported on a single crystal of Ti (see Table.~\ref{table.3}). This is due to the anisotropic elasticity, where the $c$  direction is stiffer than the $a$ direction. Therefore, as temperature increases, strain caused by thermal expansion in the $c$ direction is expected to be less than that of the $a$ direction. Hence, lower thermal expansion coefficient is expected. In comparison with literature values of single crystal Ti, thermal expansion coefficient along the $c$ direction, $\alpha_{33}$ agrees well while along the $a$ direction, $\alpha_{11}$ was found to be lower. This difference is likely from the elastic constraint of a single grain within a polycrystalline aggregate, compared to a single crystal which may expand without restriction. With the constraint in a polycrystal, $\alpha_{11}$ is found to be lower compared to the value of a single crystal. 

As the grain-grain interactions can be successfully captured in the crystal plasticity model, the thermal properties implemented in the model are the values of a single crystal of Ti shown in Table~\ref{table.3}. This can be verified by lattice strains that developed along the loading direction by increasing the temperature as shown in Fig.~\ref{fig.5}(b) and~\ref{fig.5}(c), where the model was able to reproduce thermal strains for the five selected plane families. These are shown to be consistent with the experimental expansion strains at all temperatures and for both grade 1 and grade 4 samples. 

Fig.~\ref{fig.5}(d) shows the change in stiffness as temperature increases, the stiffness values were obtained from the gradients of stress-lattice strain curves during elastic loading. Six plane families with lower noise level in stiffness were plotted. A linear trend is seen; as temperature increases the stiffness decreases. 

\subsection{Macroscopic stress relaxation}

\begin{figure}[p]
\centering
\includegraphics[width=0.875\textwidth]{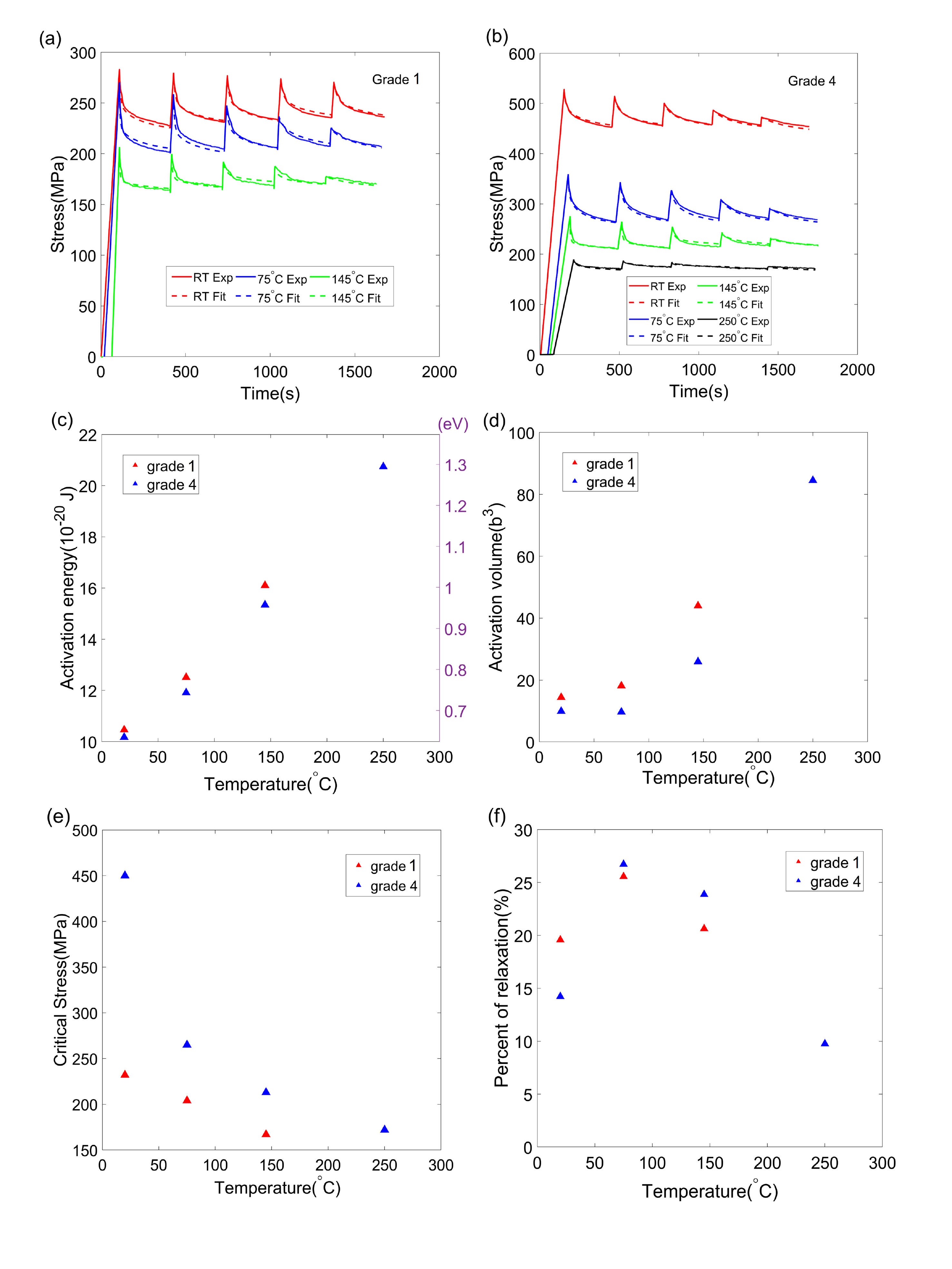}
\caption{(a) Macroscopic stress relaxation curves of CP-Ti grade 1 at 3 different temperatures with dashed lines showing a fitted function based on the slip law; (b) Macroscopic stress relaxation curves of CP-Ti grade 4 at 4 different temperatures; (c) Macroscopic thermal activation energy, $\Delta{F}$ vs. temperature; (d) Macroscopic activation volume, $\Delta{V}$ vs. temperature; (e) Macroscopic critical stress, $\sigma_c$ vs. temperature; (f) Percentage of stress relaxed vs. temperature.}
\label{fig.6}
\end{figure}

The macroscopic stress relaxation responses, calculated from the measured force from the ETMT load cell are shown in Fig.~\ref{fig.6}(a) and~\ref{fig.6}(b). They were processed using the same method as described in the previous work~\cite{XIONG2020}. Briefly, a constitutive law that links the plastic normal strain rate, $\dot{\epsilon}_p$, to the macroscopic normal stress, $\sigma$, was used:
\begin{equation}\label{eq8}
    \dot{\epsilon}_p = \rho b^2 \nu \exp\left(-\frac{\Delta{F}}{k_BT}\right)\sinh\left(\frac{(\sigma-\sigma_c) {\Delta}V}{k_BT}\right)
\end{equation}
\begin{equation}\label{eq9}
    \Delta{\epsilon}_e = -\Delta{\epsilon}_p = -\dot{\epsilon}_p\Delta{t} \rightarrow \Delta{\sigma} = E\Delta{\epsilon_e} \rightarrow \sigma_{i+1} = \sigma_i + \Delta{\sigma}
\end{equation}

The initial stress, $\sigma_i$, obtained from the experimental stress at the beginning of the first stress relaxation cycle, was substituted in to the Equation.~\ref{eq8} and yield a macroscopic plastic strain rate, $\dot{\epsilon}_p$. The change in macroscopic plastic strain, $\Delta{\epsilon}_p$, was calculated over a small time step $\Delta{t}=0.2$~s. This change in macroscopic plastic strain matches with the change in macroscopic elastic strain, $\Delta{\epsilon}_e$, as the total strain was held constant. Therefore, the change of macroscopic stress (used for update the stress of the next time step), $\Delta{\sigma}$, was obtained by the product of the Young’s modulus \textit{E} for Ti at each temperatures~\cite{ZHANG2015,STRINGER1960} and $\Delta{\epsilon}_e$. This process was repeated to reconstruct the stress relaxation curves. With the aid of fitting tools in Matlab, the best-fit relaxation curve (dashed lines in Fig.~\ref{fig.6}(a) and~\ref{fig.6}(b)) was found with fitting error less than 1\% and the optimal fitting parameters $\Delta{F}^{macro}$,$\Delta{V}^{macro}$ and $\sigma_c^{macro}$ can be obtained, which are shown in Fig.~\ref{fig.6}(c),~\ref{fig.6}(d) and~\ref{fig.6}(e).

Both activation energy and activation volume were found to increase with temperature; the activation energy has a linear response with temperature, which replicates observations in BCC steel~\cite{TANAKA2016}. Grade 1 has a higher activation energy than that of grade 4, indicating higher thermal energy barrier for grade 1 over grade 4. The activation volume was also found to be higher in grade 1 over grade 4, because grade 1 has lower oxygen content than grade 4. Oxygen is known to be an interstitial in $\alpha$-Ti and has a profound strengthening effect by acting as obstacles for dislocation motion~\cite{DUNNE2008,CONRAD1981}. As the activation volume is proportional to the pinning distance, which is inverse proportional to obstacle density~\cite{DUNNE20071061}, a higher activation volume is expected in a lower obstacle density material, therefore was observed for grade 1 CP-Ti. The normal critical stress, known as the stress for the onset of plasticity, decreases with temperature and for grade 4, was found to be significantly higher than that of grade 1, due to the strengthening effect of oxygen~\cite{CONRAD1981,BRITTON2015,CHURCHMAN1954,GUO2014}.  

During the five-minute stress relaxation period, the total strain was held constant and due to the time dependent plasticity, there was an increment of plastic strain and a relaxation of elastic strain and stress. Therefore, the amount of stress relaxed during the strain hold reveals the amount of plasticity and thus the activity of dislocations. As each sample was loaded to different macroscopic stresses, direct comparison of the absolute magnitude of the relaxed stress is not possible. However, ratios of relaxed stress to the initial stress at the beginning of the first stress relaxation cycles (percentage of relaxation) were calculated for a better comparison. Fig.~\ref{fig.6}(f) shows the variation of percentage relaxation versus temperatures, where in both grade 1 and grade 4, the percentage relaxation first increases when the temperature increases to 75$^{\circ}$C and then decreases as temperature further increases. A peak value is expected around 75$^{\circ}$C. At room temperature, dislocations in grade 1 are more active than in grade 4, but at higher temperatures (75$^{\circ}$C and higher) the activity of dislocation in grade 4 exceeds grade 1 results from a sharper increase in percentage of relaxation in grade 4.  

\subsection{Lattice strain simulation results}
The change in mobile dislocation density is small during each stress relaxation period~\cite{WANG2006,XIONG2020}, which is verified by fitting the macroscopic stress relaxation cycles with a function informed by a slip law (as discussed in section 3.2), where one set of parameters can be fitted to all five relaxation cycles. Therefore, the lattice strains developed during the first 5-minute strain hold from the five selected planes families were used to calibrate the crystal plasticity model. Because the loading strain rate is relatively slow (\textit{ca.} $3\times10^{-5}~\text{s}^{-1}$) in this work, at this strain rate the lattice strains are less sensitive to the change in the activation volume, $\Delta{V}$. The activation volume used in the simulation are the same as those obtained from the macroscopic stress relaxation fit. Considering the activity of the pyramidal slip is low, parameters for this slip type was not calibrated. The macroscopic $\Delta{F}$ and CRSS were instead set to be three times the value of basal slip~\cite{XIONG2020}. Efforts were focused on the identification of the parameters: activation energy and CRSS for (1) basal and (2) prism slip to enable simulation of the lattice strains from relaxation cycles, which could then be fitted to the experimentally measured lattice strains.

Among the five selected plane families, three important planes in the HCP crystal structure (prismatic planes \{01$\bar{1}$0\}, first ordered pyramidal planes \{01$\bar{1}$1\} and second ordered pyramidal planes \{11$\bar{2}$2\}) have a higher priority in the optimisation process. Fig.~\ref{fig.7} shows a comparison of the simulated and experimental lattice strain relaxation curves, all of them show a good fit. Meanwhile, as the macroscopic stress relaxation curves also show good agreement with the simulated and experimental results (as shown in Fig.~\ref{fig.8}), the confidence in the measurement of the parameters is high. 
\begin{figure}[b!]
\centering
\subfigure{
\includegraphics[width=1\textwidth]{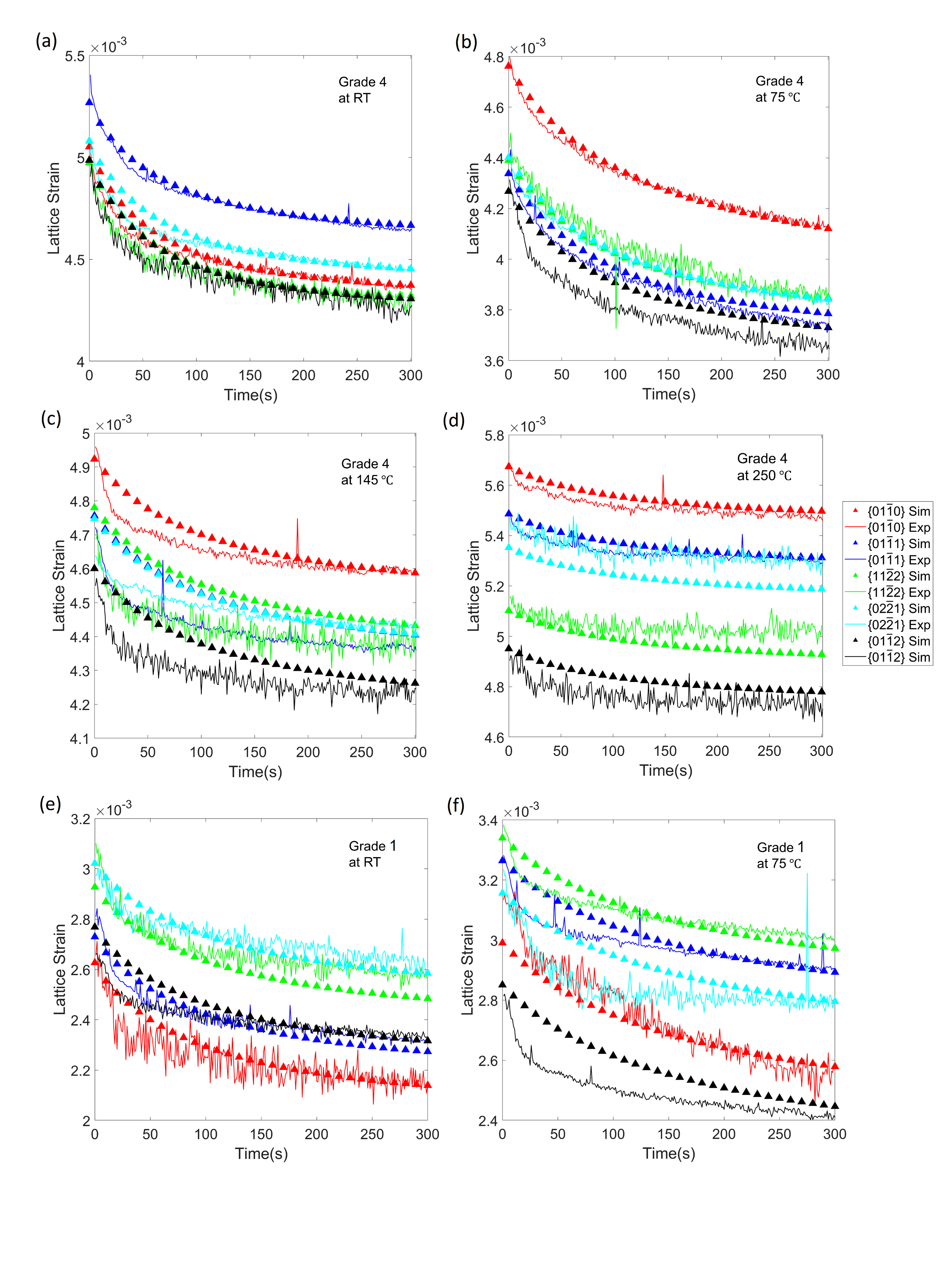}
}
\phantomcaption
\end{figure}
\begin{figure}[t!]\ContinuedFloat
\centering
\subfigure{
\includegraphics[width=.55\textwidth]{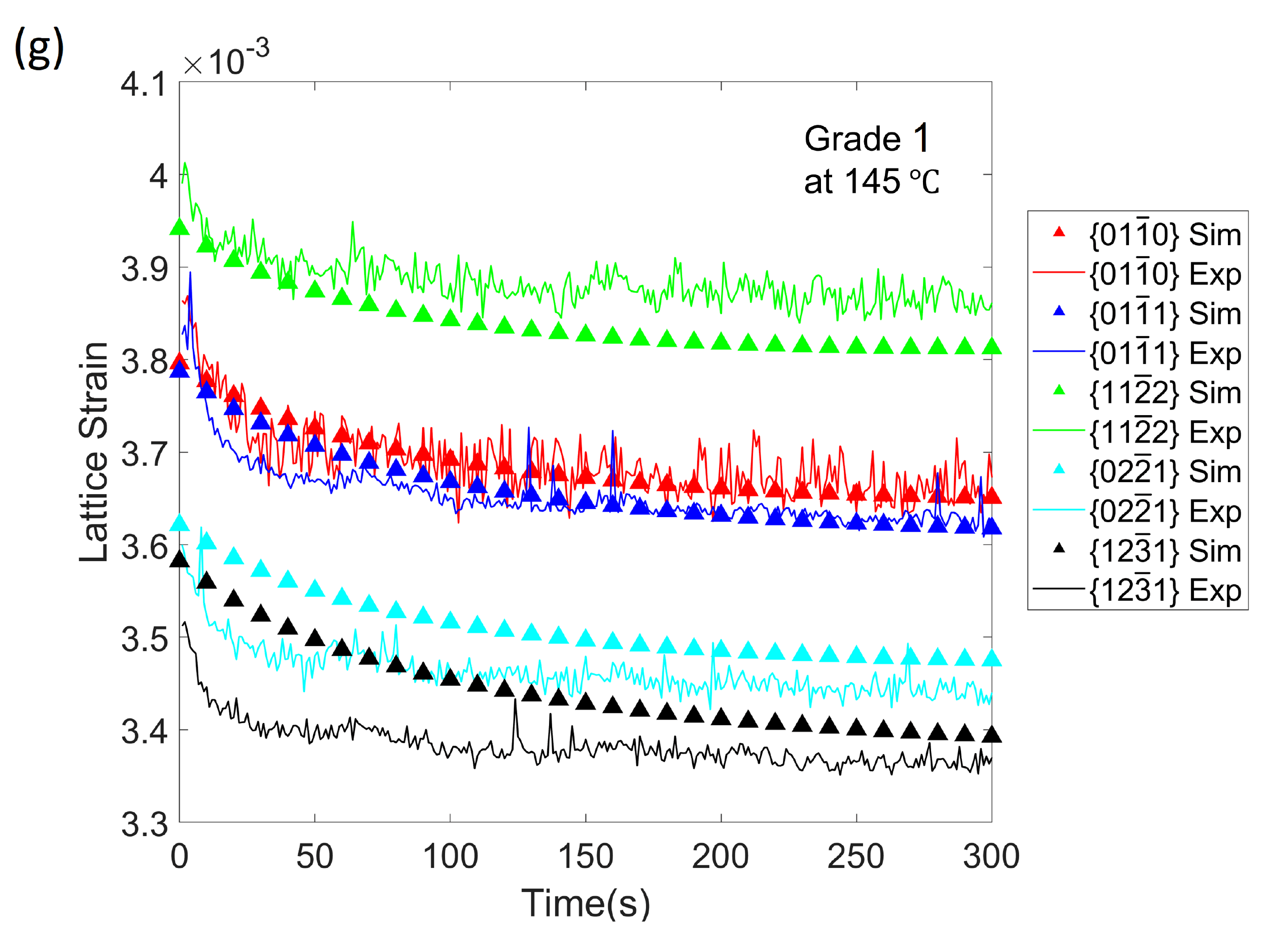}
}
\caption{Comparison of lattice strain evolution along the loading direction between the experiment and the simulation of (a) CP-Ti grade 4 at room temperature; (b) CP-Ti grade 4 at 75~$^{\circ}$C; (c) CP-Ti grade 4 at 145~$^{\circ}$C; (d) CP-Ti grade 4 at 250~$^{\circ}$C; (e) CP-Ti grade 1 at room temperature; (f) CP-Ti grade 1 at 75~$^{\circ}$C; (g) CP-Ti grade 1 at 145~$^{\circ}$C.(Note vertical axes are not the same in each plot)}
\label{fig.7}
\end{figure}
\begin{figure}[hbt!]
\centering
\includegraphics[width=1\textwidth]{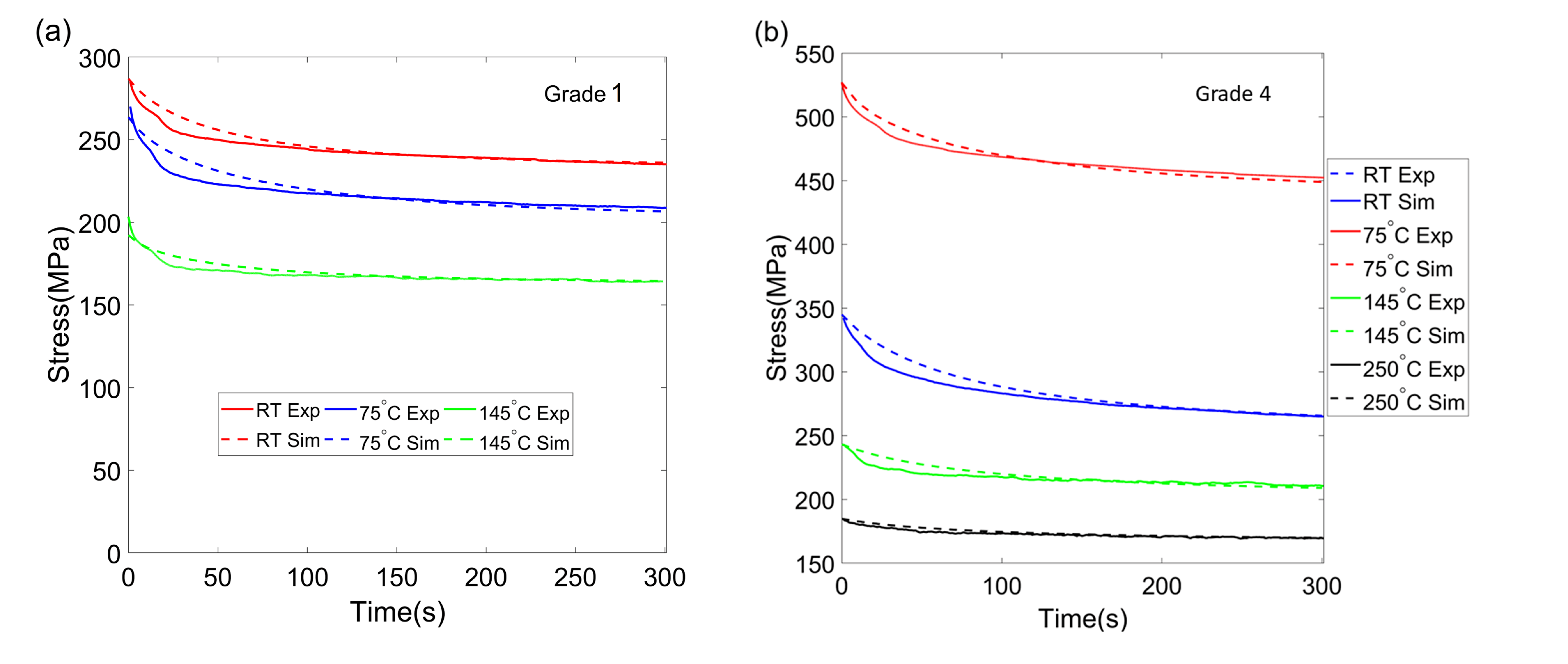}
\caption{Comparison of macroscopic stress relaxation curves between experiment and simulation for (a) CP-Ti grade 1 at 3 different temperatures and (b) CP-Ti grade 4 at 4 different temperatures.}
\label{fig.8}
\end{figure}
Fig.~\ref{fig.9}(a) and~\ref{fig.9}(b) show the variation of CRSS and activation energy for the basal and prism slip as a function of temperature. Error bars were shown to give the range of these parameters i.e. changing parameters within these ranges results in a change in simulated lattice strain that is smaller than the experimental results error (fluctuation), and simulated lattice strain remains good fit with experiment lattice strain. 

CRSS values decrease as temperature increases. At room temperature, the CRSS for basal slip is significantly higher than that of prism slip in both grade 1 and grade 4. As the temperature increases, the CRSS for basal slip drops much faster than that of prism slip in grade 4. This results in CRSS value for basal slip to be lower than that of prism slip in CP-Ti grade 4 at 250~$^{\circ}$C. However, this trend was not found in grade 1, where the difference between values of basal and prism slip remains almost the same for all three temperatures. Due to the strengthening effect of oxygen, grade 4, with higher oxygen content, has a higher CRSS for prism slip at all temperatures and basal slip at room temperature over grade 1. However, the CRSS for basal slip at 75~$^{\circ}$C and 145~$^{\circ}$C were found to be similar for these two CP-Ti grades. As a result, the ratio of $\tau_c^{basal}$ and $\tau_c^{prism}$ was found to be higher in grade 1 over grade 4 at room temperature, as temperature increases, the CRSS ratio was found to decrease in grade 4 while it increases in grade 1.

This phenomenon can be explained by the difference in the strengthening effect of oxygen on these two slip systems. Yu \emph{et al}~\cite{Yu2015} suggested that there is a very strong repulsive interaction between a screw dislocation core and oxygen on octahedral site on the same prismatic plane. On the contrary, the repulsion of a dislocation core and oxygen on an interstitial site on basal plane is an order of magnitude smaller. It was also found that once interacting with dislocation, the original oxygen octahedral site will gradually disappear. The oxygen would migrate to either a basal plane interstitial site (with lower energy) or a tetrahedral site and a new octahedral site on a prism plane (with higher energy). As the temperature increases, the thermal expansion in the $a$ direction is larger than that of the $c$ direction (as discussed in section 3.1), which results in a decrease in the $c/a$ ratio. Therefore, the area of the basal plane becomes larger ($a_1$ and $a_2$ vectors increase, as $c$ decreases) and the octahedral site becomes nearer to the basal plane, which results in a higher possibility for oxygen to move to a basal-plane interstitial sites. For a lower oxygen content Ti (grade 1); this results in a strengthening effect loss for the prism type slip. For a higher oxygen content Ti (grade 4), although some of the oxygen may migrate to a different interstitial site, the remaining oxygen still provides a much more significant strengthening effect for prism slip. To summarise the temperature effect and oxygen level effect, we can correlate the observation of the changing in CRSS ratio vs. temperatures have opposite trend in two grades of CP-Ti with different oxygen content.      

The thermal activation energy shows an opposite trend to CRSS, which increases with temperature (as shown in Fig.~\ref{fig.9}(b)), which is a trend commonly observed in coarse-grained materials~\cite{TANAKA2016}. It is found that the thermal activation energy for both basal and prism slip in grade 1 is higher than that in grade 4, indicating higher thermal energy barrier reduces with increasing oxygen content. Basal slip generally has a higher thermal energy barrier over prism slip except for grade 4 at 250~$^{\circ}$C. 
\begin{figure}[hbt!]
\centering
\subfigure{
\includegraphics[width=.46\textwidth]{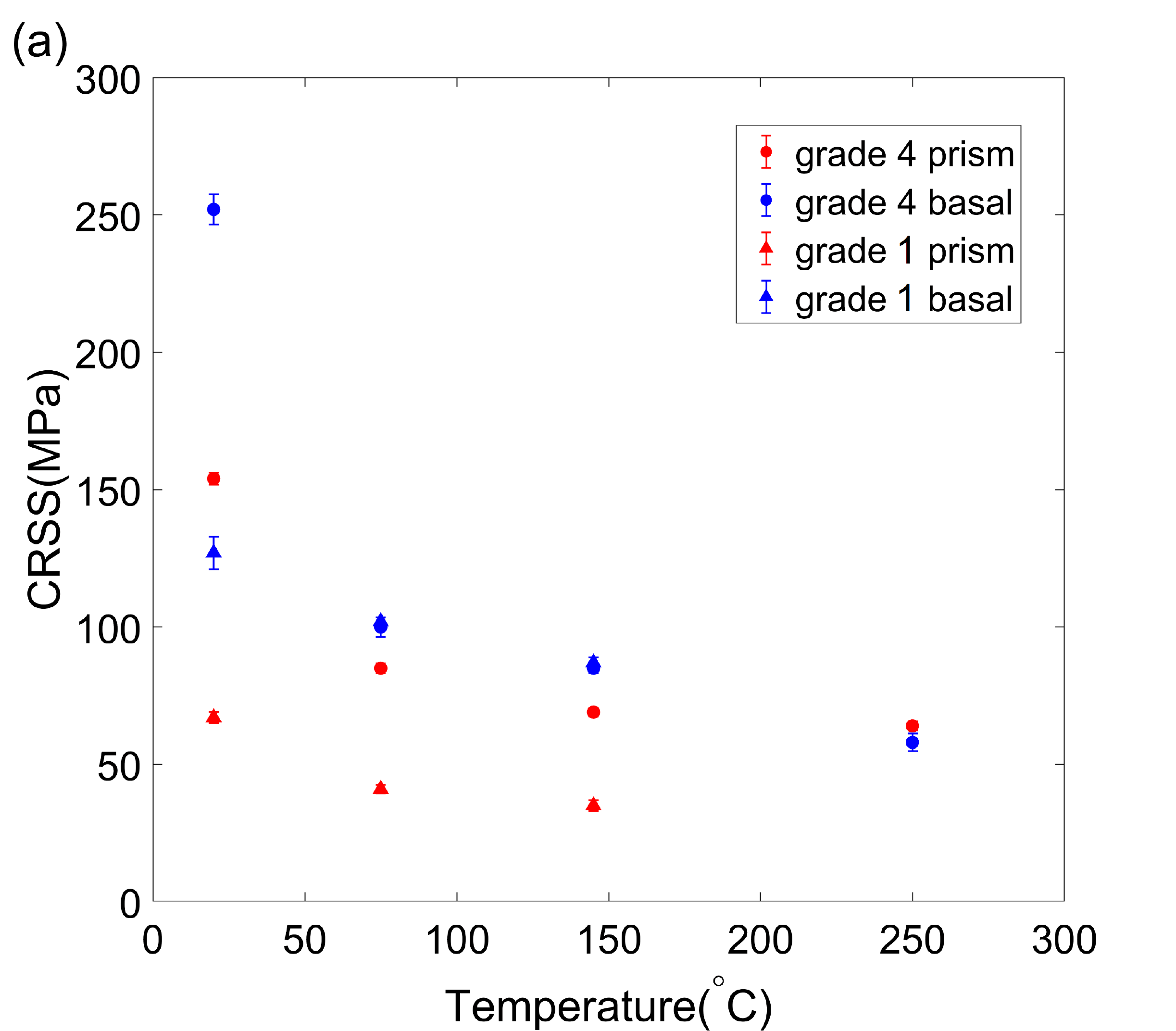}
}
\subfigure{
\includegraphics[width=.45\textwidth]{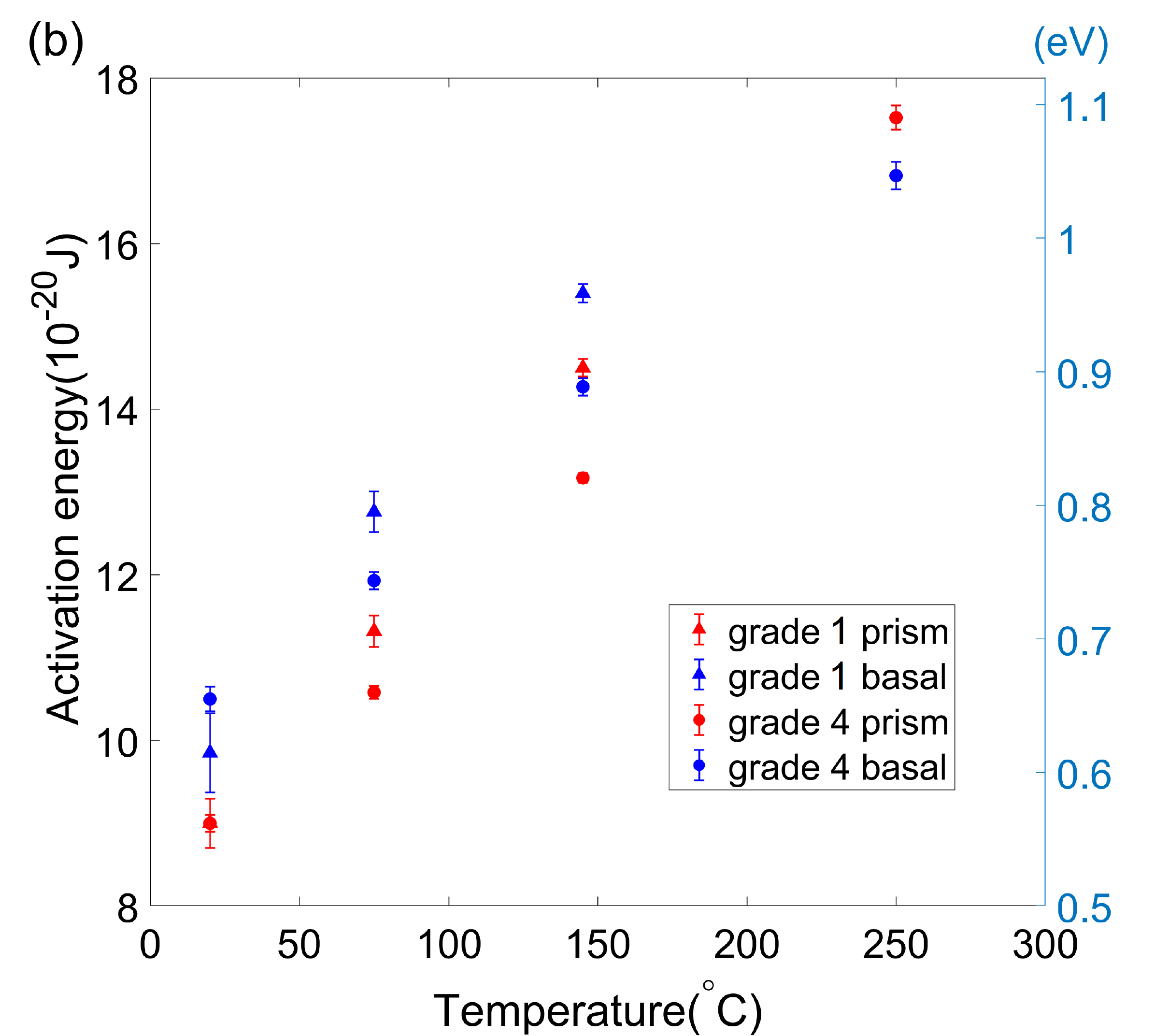}
}
\subfigure{
\includegraphics[width=.46\textwidth]{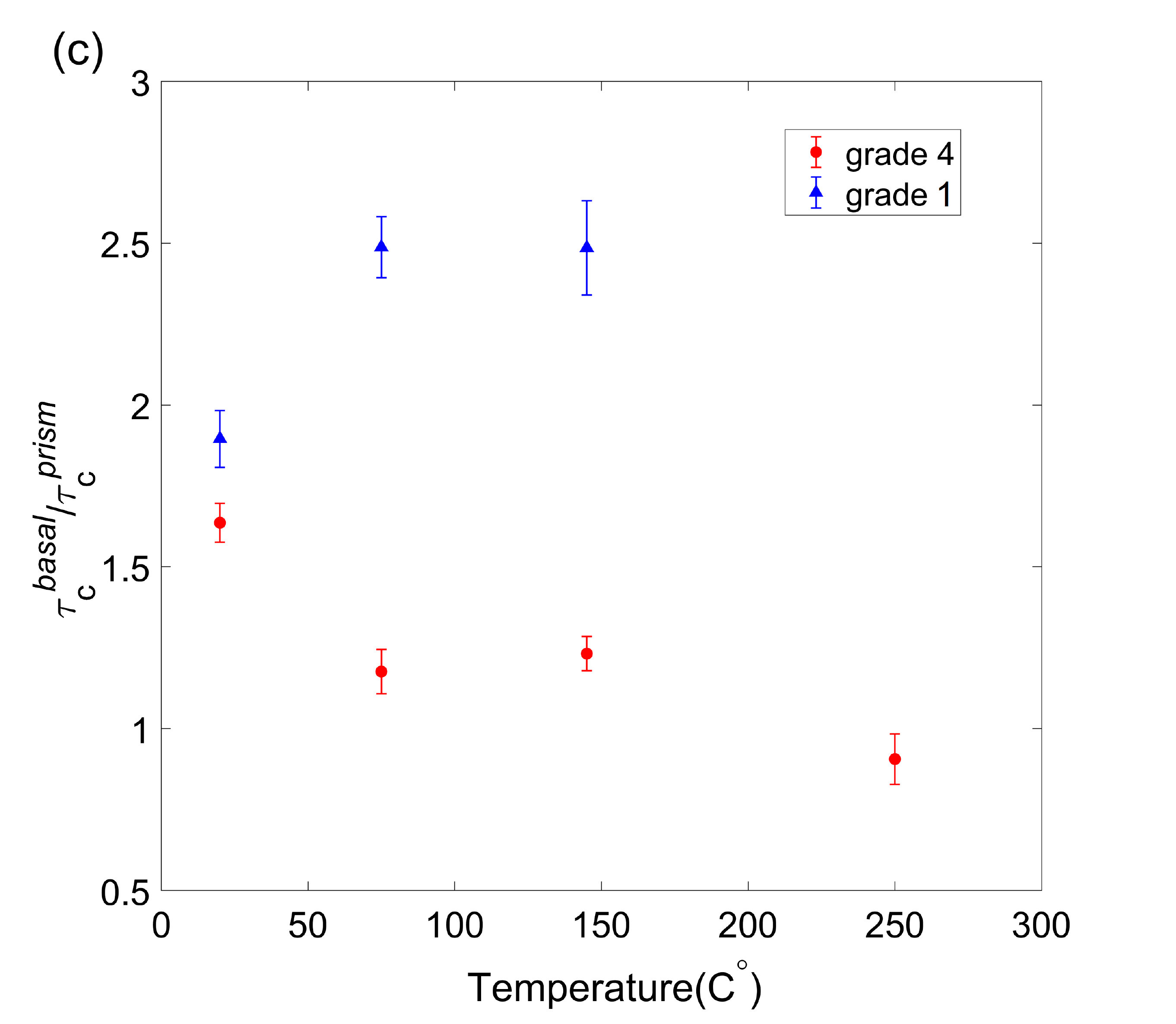}
}
\subfigure{
\includegraphics[width=.46\textwidth]{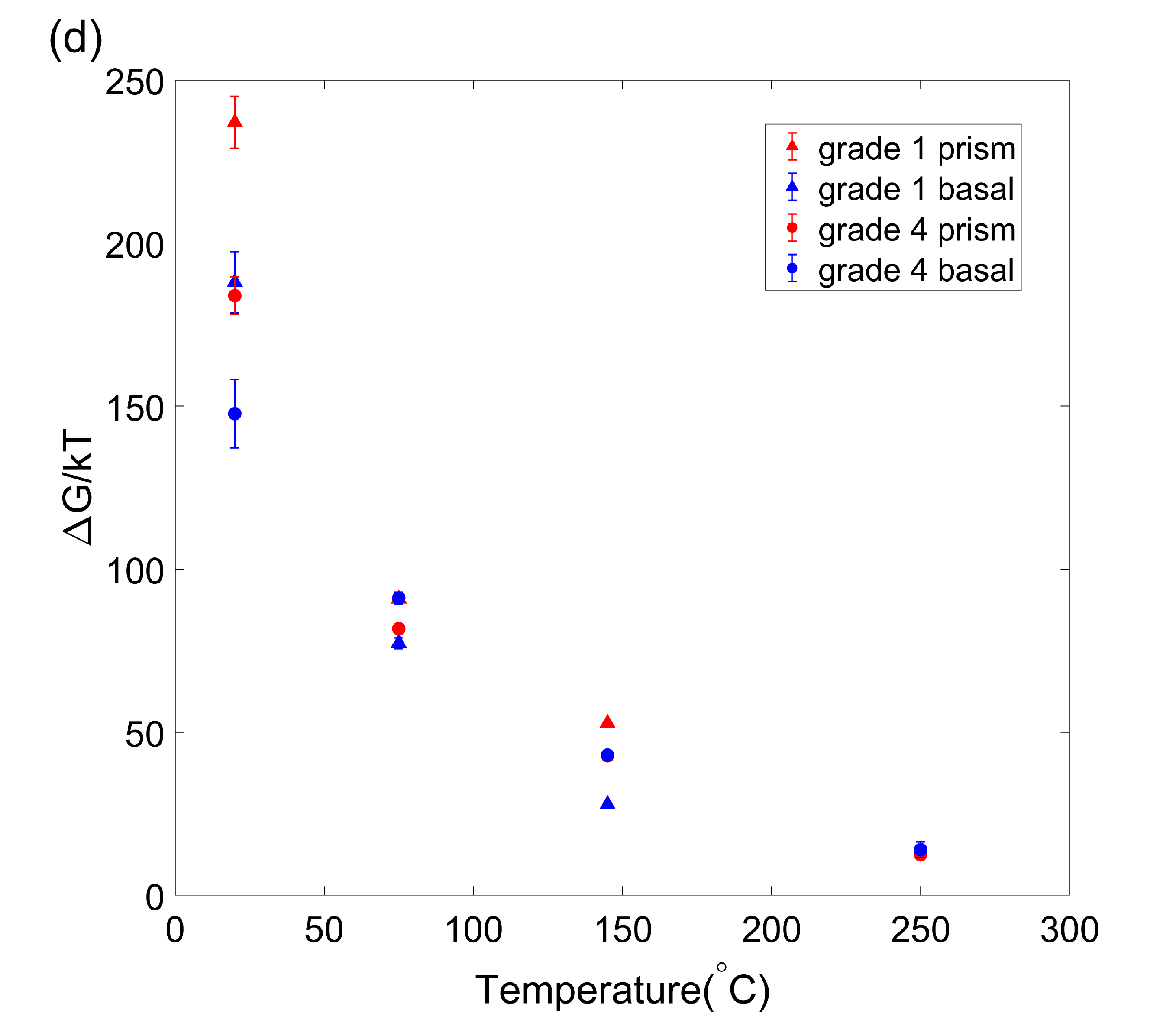}
}
\caption{(a) CRSS vs. temperature; (b) Thermal activation energy, $\Delta{F}$ vs. temperature; (c) CRSS ratio vs. temperature; (d) Activation free energy as a multiple of available thermal energy, $\Delta{G}/k_BT$ vs. temperature.}
\label{fig.9}
\end{figure}

The total activation free energy barrier has two contributing factors: (1) the thermal activation energy barrier, $\Delta{F}$ and (2) the critical mechanical work done by an externally applied stress, $\tau_c\Delta{V}$. Therefore, the total free energy barrier for dislocation activation is given by~\cite{DUNNE20071061}:
\begin{equation}\label{eq10}
    \Delta{G} = \Delta{F} - \tau_c\Delta{V}
\end{equation}
where $\Delta{G}$ is the Gibbs free energy. This expression was then normalised by absolute temperature, $T$ and Boltzmann constant, $k_B$ giving: $\Delta{G}/k_BT$. Fig.~\ref{fig.9}(d) shows the variation of the energy barrier for dislocation activation against temperature; it is found that as the temperature increases, $\Delta{G}/k_BT$ for both slips systems in grade 1 and grade 4 decreases, indicating that slip becomes easier to activate as temperature increases. Meanwhile, due to the reverse response of $\Delta{F}$ and $\tau_c$  as the temperature is increased, the slip activation mechanism changes, where at low temperature slip activation is mainly driven by mechanical work while at high temperature slip activation is more driven by thermal energy. 
\subsection{Strain rate sensitivity}
\begin{figure}[h!]
\centering
\subfigure{
\includegraphics[width=.47\textwidth]{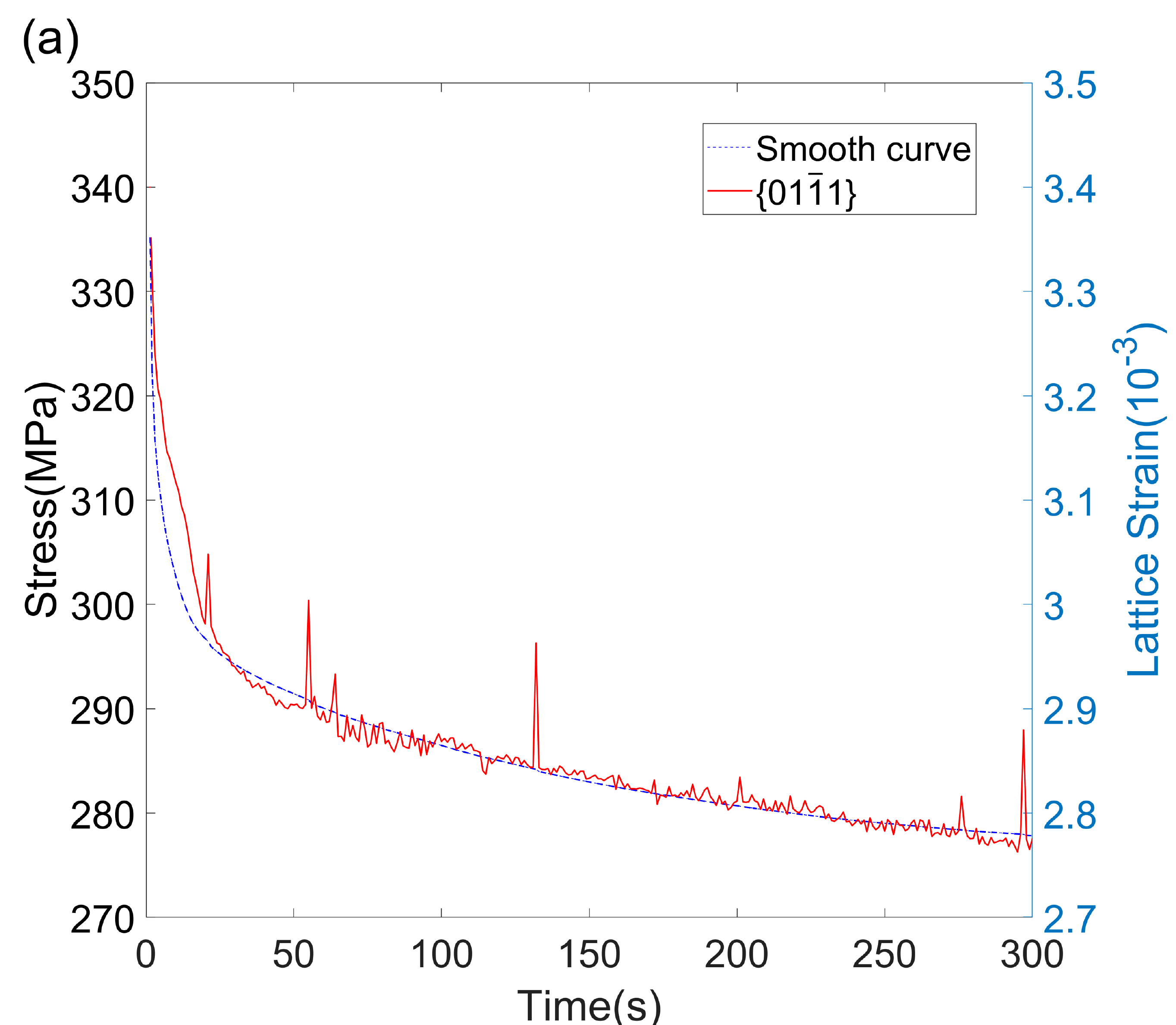}
}
\subfigure{
\includegraphics[width=.44\textwidth]{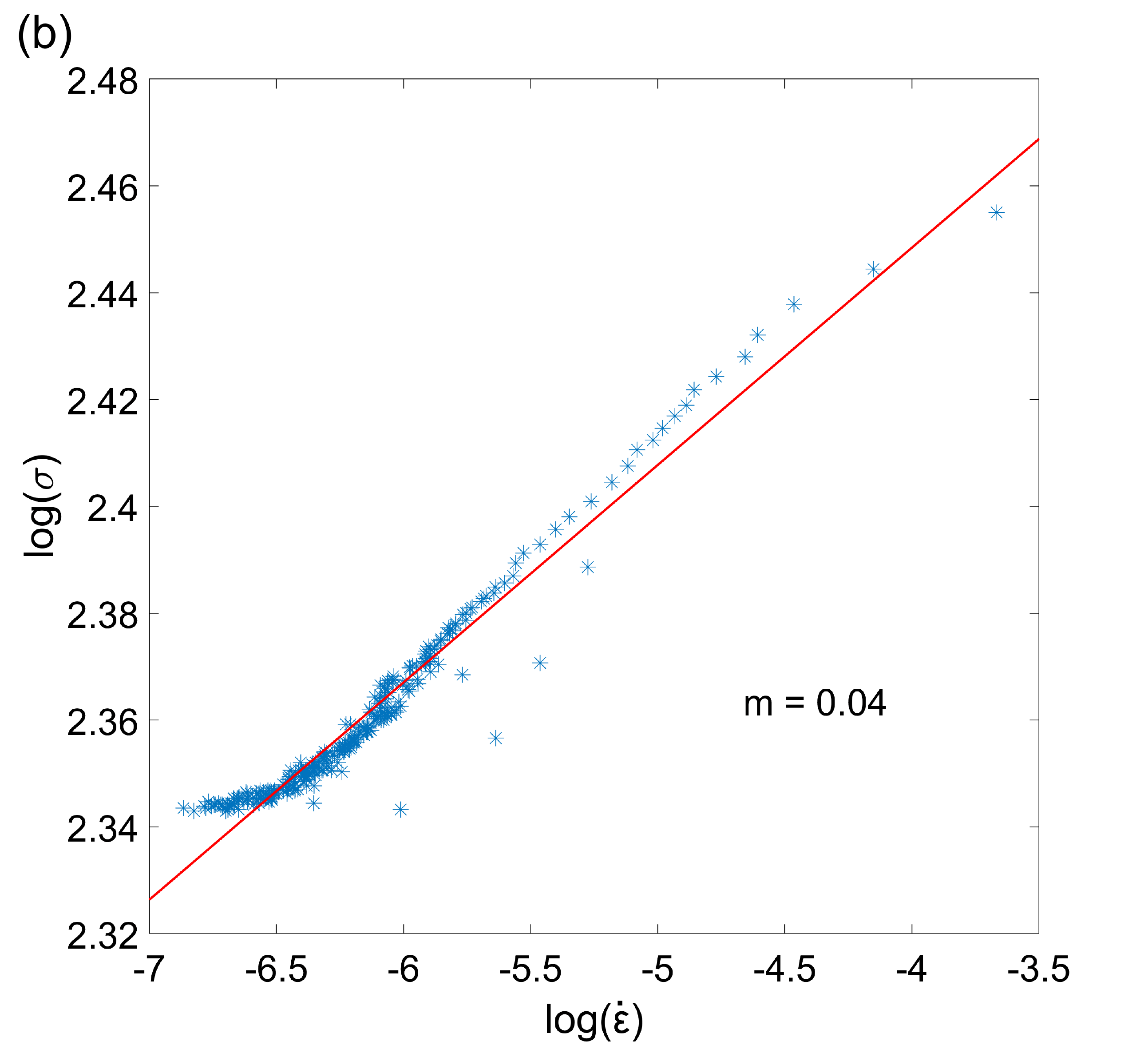}
}
\caption{(a) Smooth fit the first stress/strain relaxation curve (stress was obtained from the product of lattice strain and plane stiffness); (b) Determination of strain rate sensitivity.(Using \{01$\bar{1}$1\} plane of CP-Ti grade 1 at 75~$^{\circ}$C as an example)}
\label{fig.10}
\end{figure}

The strain rate sensitivity exponent, $m$, can be calculated follow the equation~\cite{JUN2016}:
\begin{equation}\label{eq11}
    m = d(log({\sigma}))/d(log({\dot{\epsilon}}))
\end{equation}
In this stress relaxation experiment, the $m$ values are obtained by the gradients of stress-strain rate plots at log scale. To do this, lattice strain during the first relaxation cycles was multiplied by the stiffness obtained in section 3.1 (Fig.~\ref{fig.5}(d)), yielding stress relaxation curves for all 21 plane families (as shown in Fig.~\ref{fig.10}(a)). The strain rate (gradient of the lattice strain relaxation curve) at each second were calculated for this 300~s period of time, so that the stress vs. strain rate can be plotted at log scale (as shown in Fig.~\ref{fig.10}(b)).  
The strain rate sensitivity therefore can be calculated for the 21 plane families in the two CP-Ti alloys. As shown in Fig.~\ref{fig.11}(a) and~\ref{fig.11}(e), the SRS values are highly anisotropic at room temperature, as the grain orientation changes from ‘hard’ ($c$-axis parallel to loading direction, declination angle is $0^{\circ}$) to ‘soft’ ($c$-axis perpendicular to loading direction, declination angle is $90^{\circ}$), SRS increases from 0 to 0.04 in CP-Ti grade 4 and to 0.035 in CP-Ti grade 1, which strongly agrees with other works on CP-Ti~\cite{JUN2016NANO,Peykov2012,CHICHILI1998,MEYERS1994,REEDHILL1995}. However, in the Ti6242 alloy, this trend was observed to be opposite by other researchers~\cite{JUN2016}. A higher SRS ($m$=0.039) was found for ‘hard’ grains over SRS ($m$=0.025) for ‘soft’ grains in $\alpha$ phase. For higher temperature samples, as shown in Fig.~\ref{fig.11}(b)-\ref{fig.11}(d),~\ref{fig.11}(f) and~\ref{fig.11}(g), the variation of SRS with orientations still exists but the SRS values are different at these temperatures. Fig.~\ref{fig.11}(h) shows the averaged SRS values for the 21 plane families at different temperatures. It was found that for both grade 1 and grade 4, averaged SRS value increases as temperature increase to 75~$^{\circ}$C and then decreases as the temperature further increases. These follow the same temperature dependence as the relaxed stress (recall Fig.~\ref{fig.6}(f)), indicating that a higher SRS leads to more plastic strain accumulation during the same time period of creep, and thus more dislocations activities.
\begin{figure}[b!]
\centering
\subfigure{
\includegraphics[width=.44\textwidth]{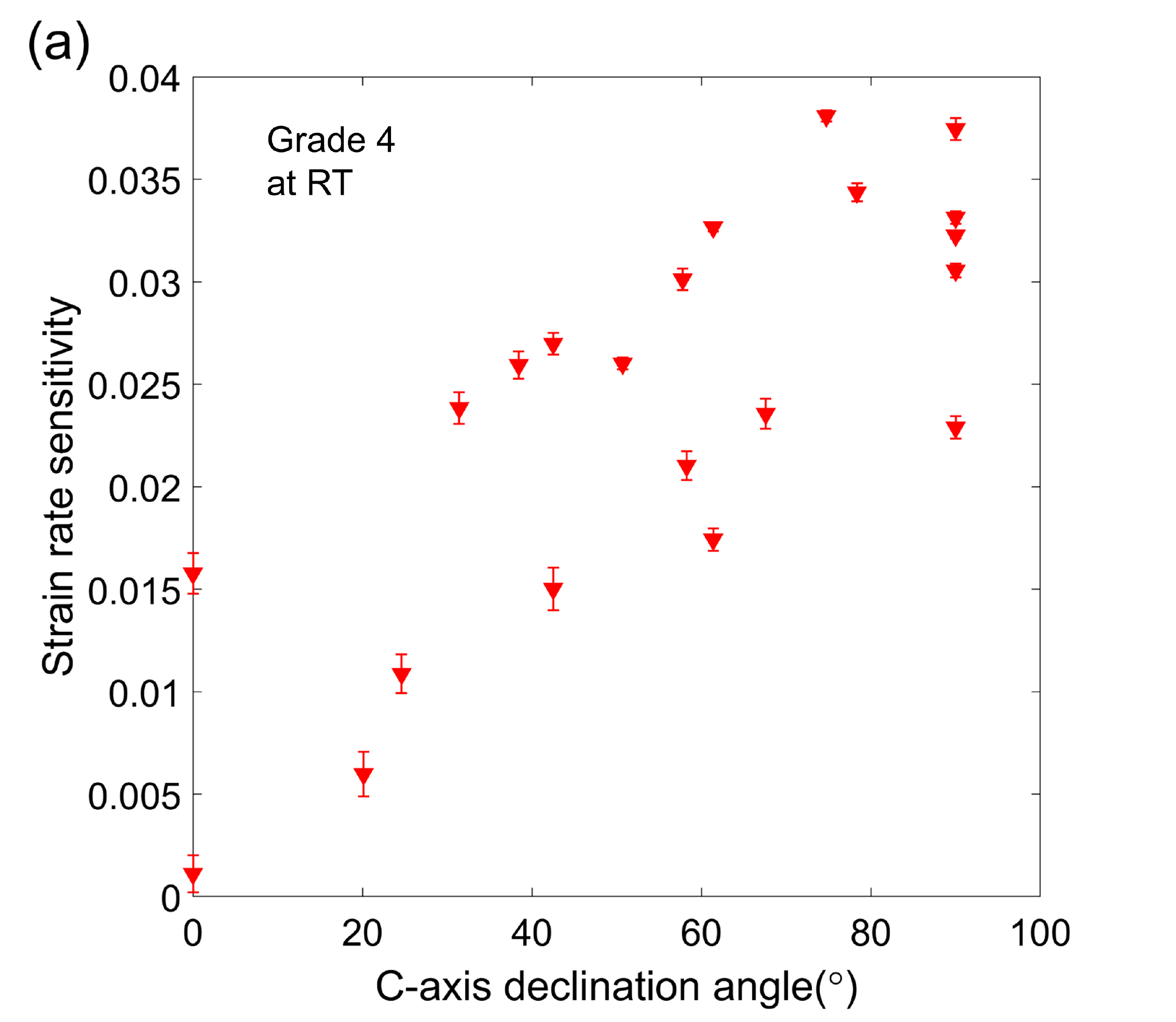}
}
\subfigure{
\includegraphics[width=.45\textwidth]{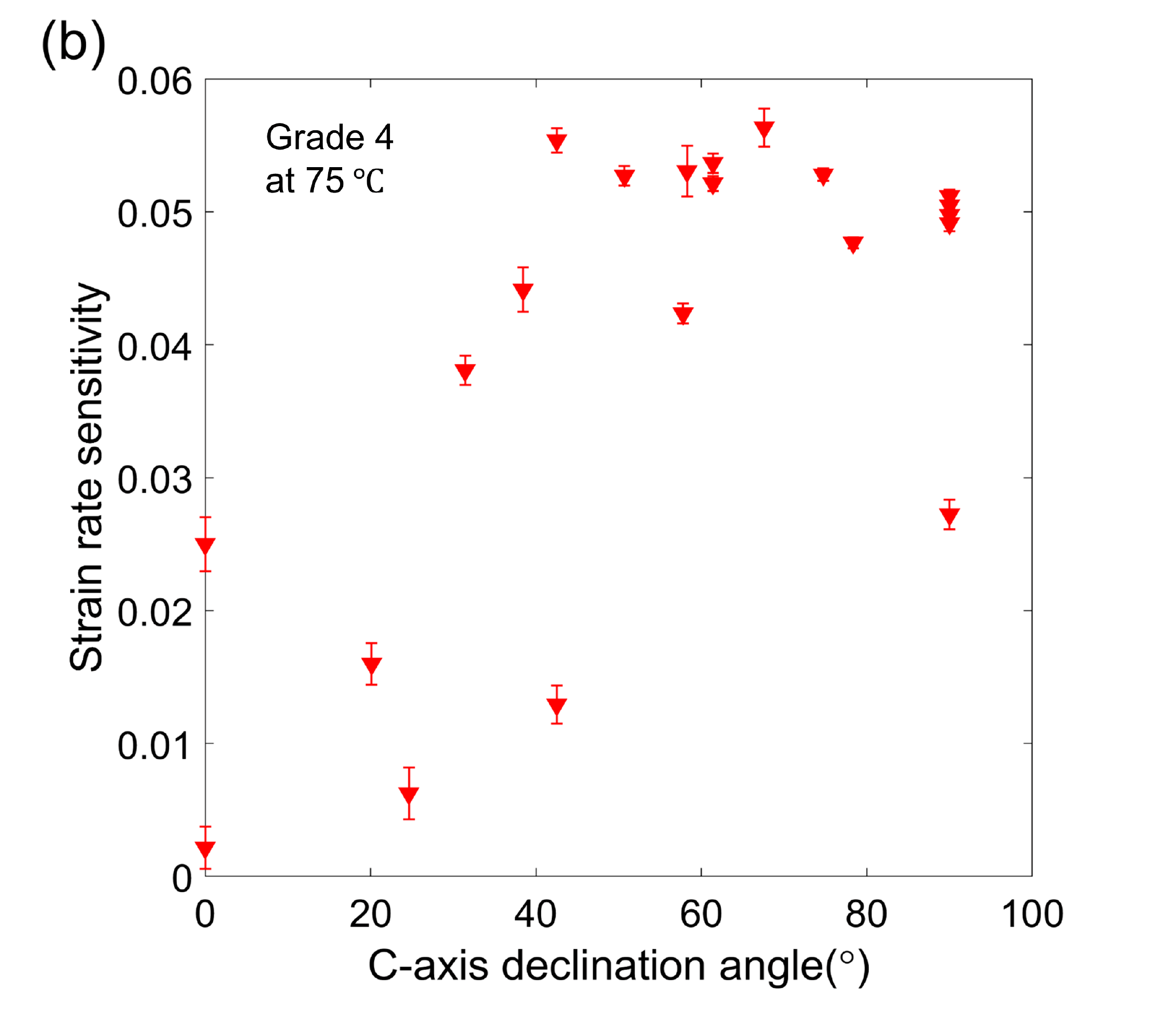}
}
\subfigure{
\includegraphics[width=.44\textwidth]{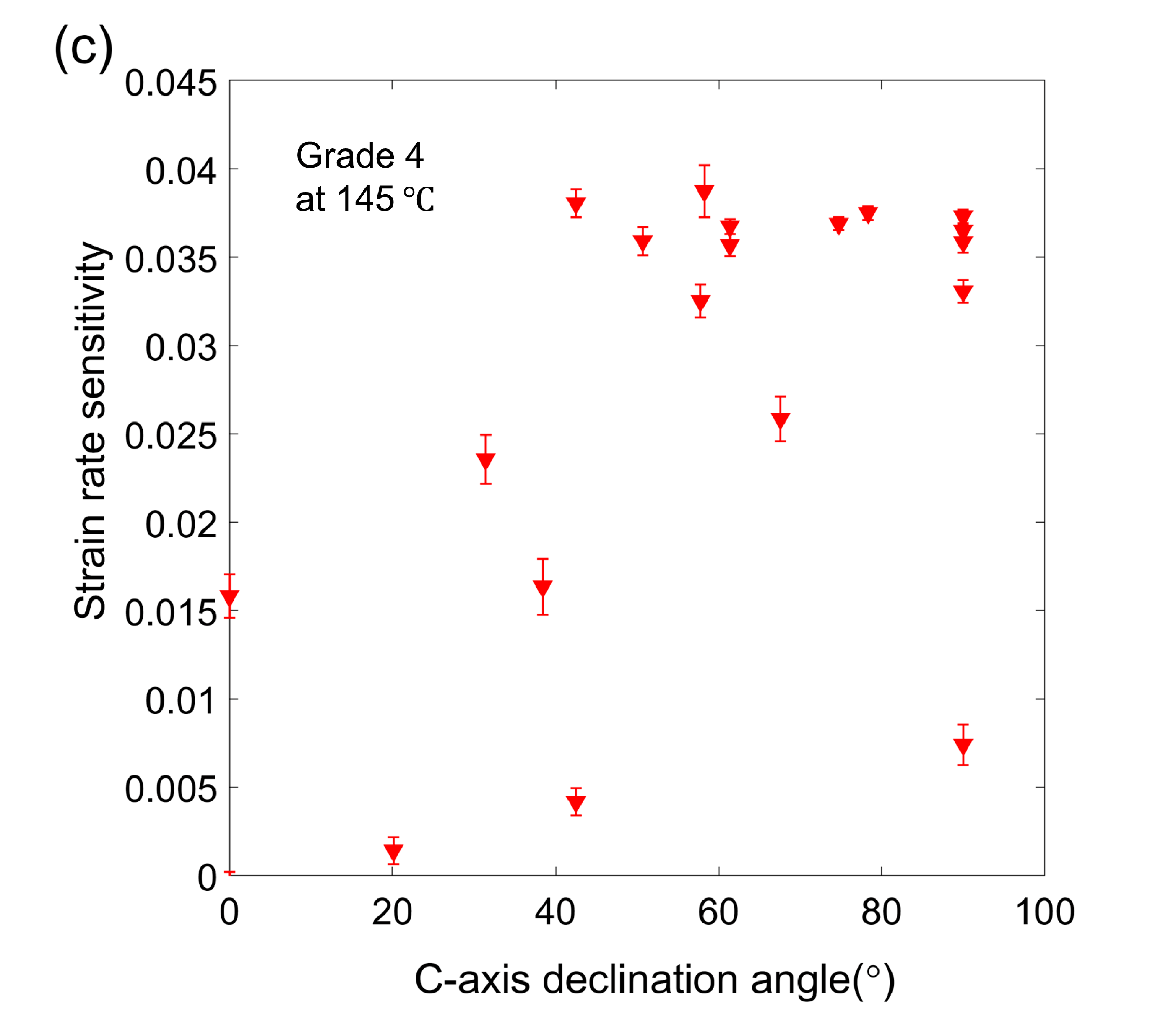}
}
\subfigure{
\includegraphics[width=.44\textwidth]{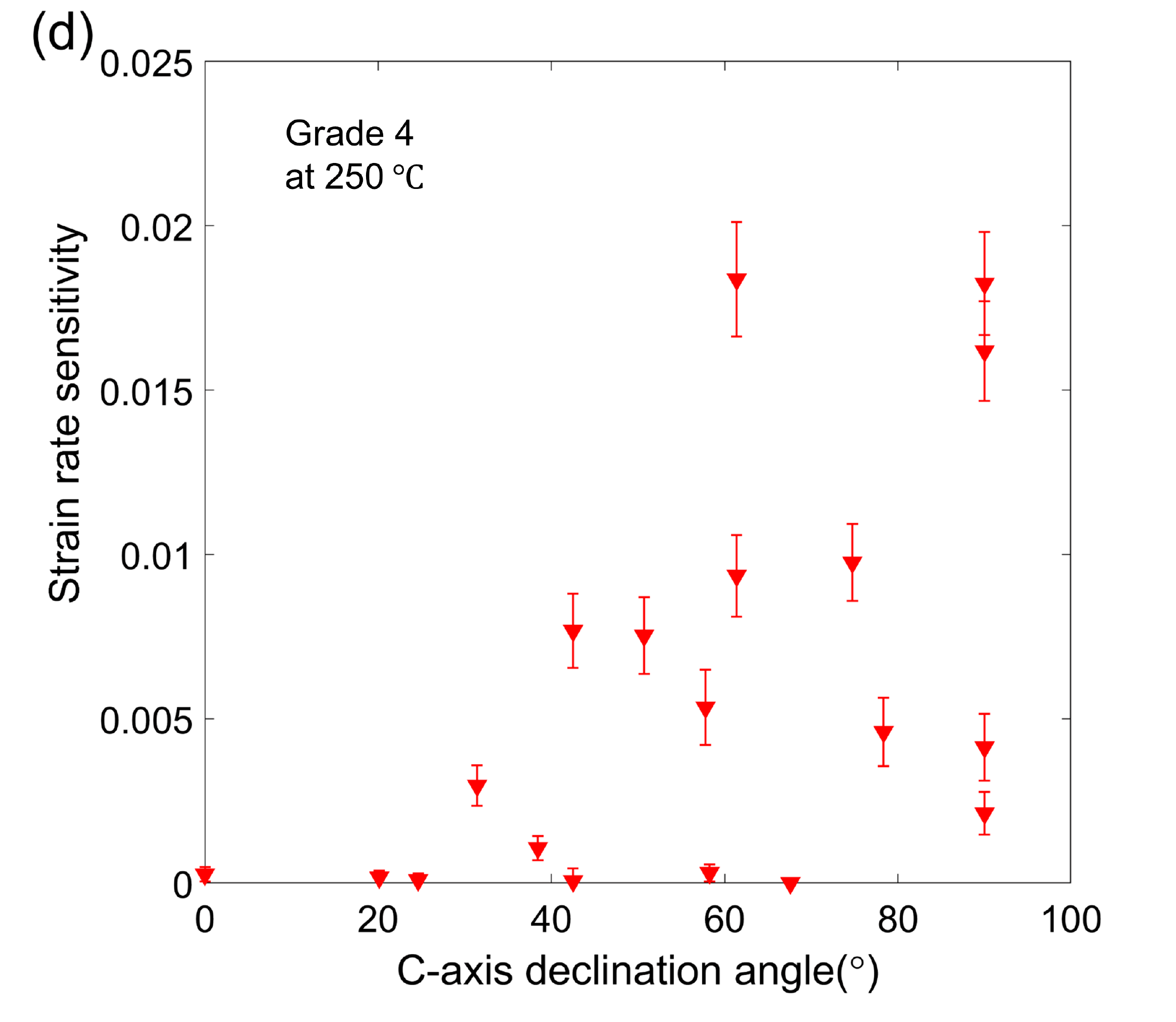}
}
\subfigure{
\includegraphics[width=.44\textwidth]{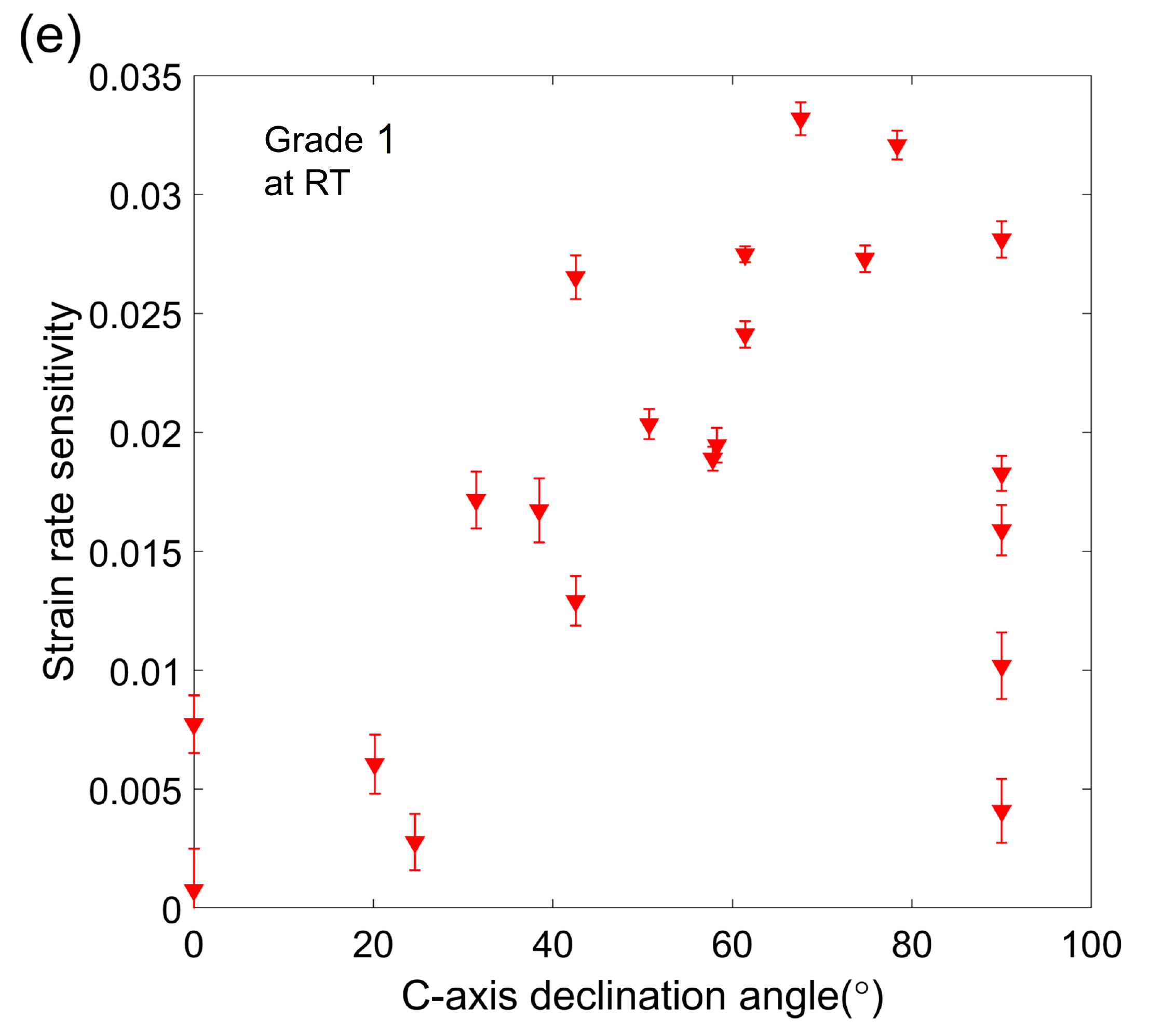}
}
\subfigure{
\includegraphics[width=.44\textwidth]{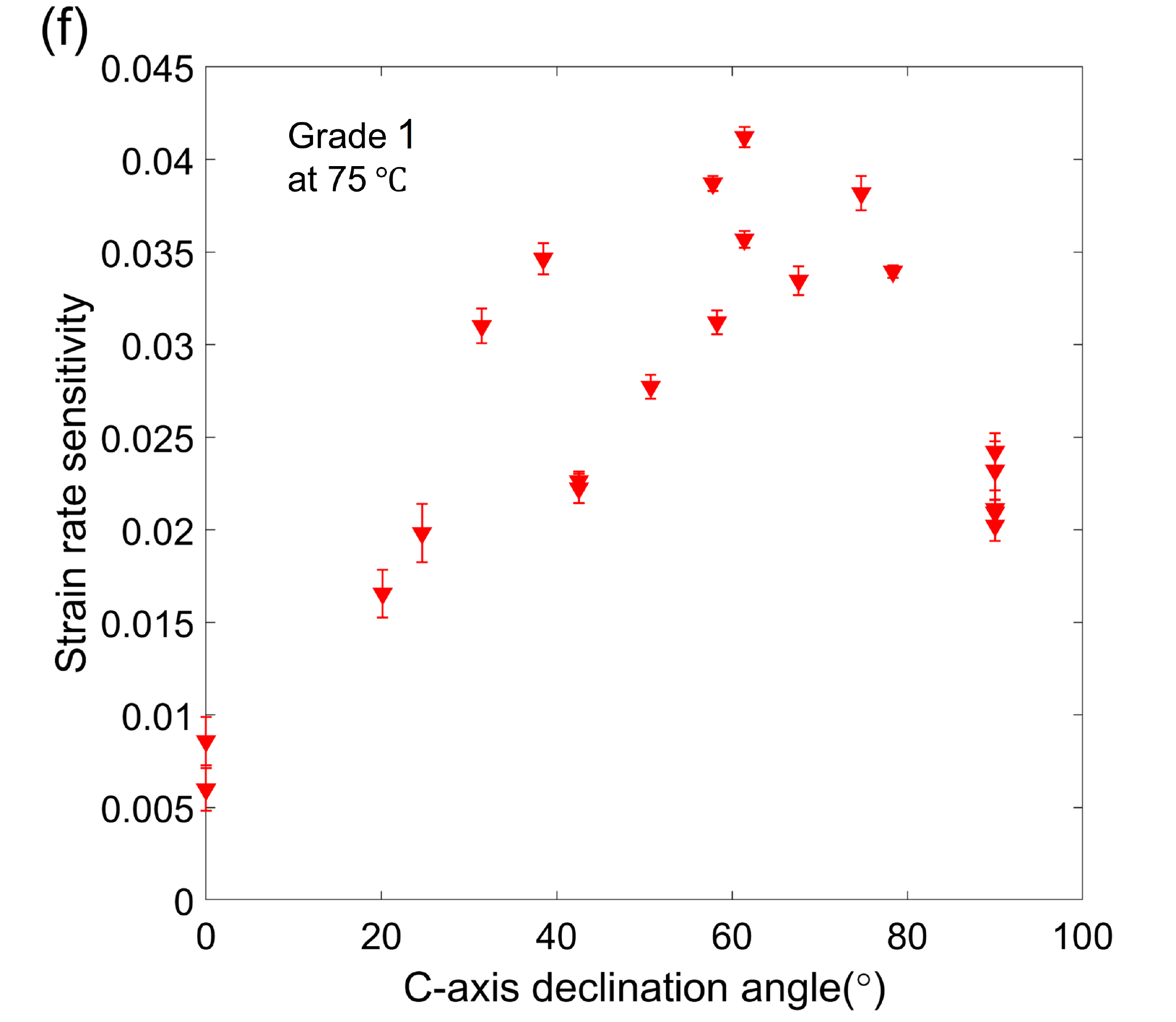}
}
\phantomcaption
\end{figure}

\begin{figure}[t!]\ContinuedFloat
\centering
\subfigure{
\includegraphics[width=.445\textwidth]{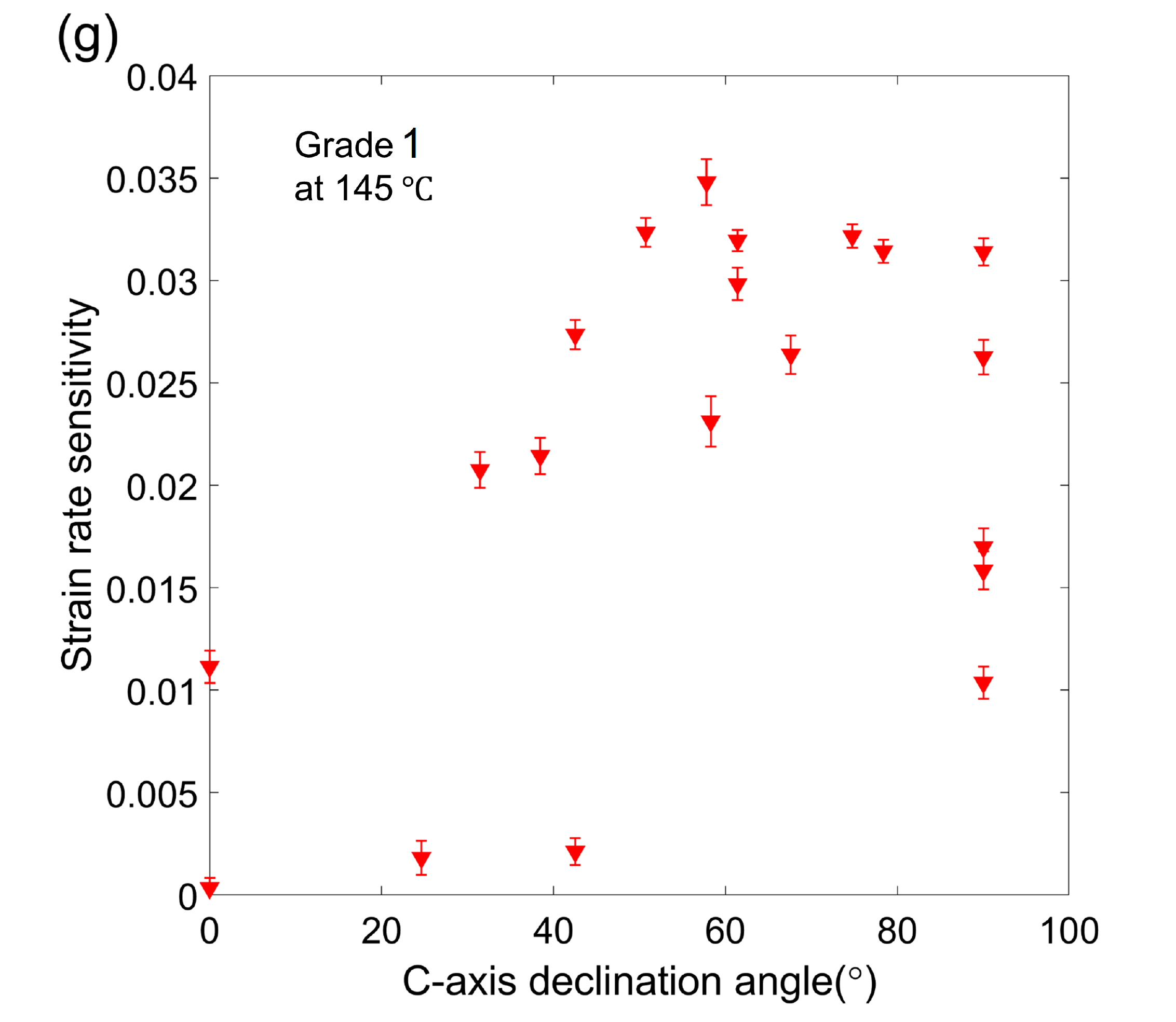}
}
\subfigure{
\includegraphics[width=.435\textwidth]{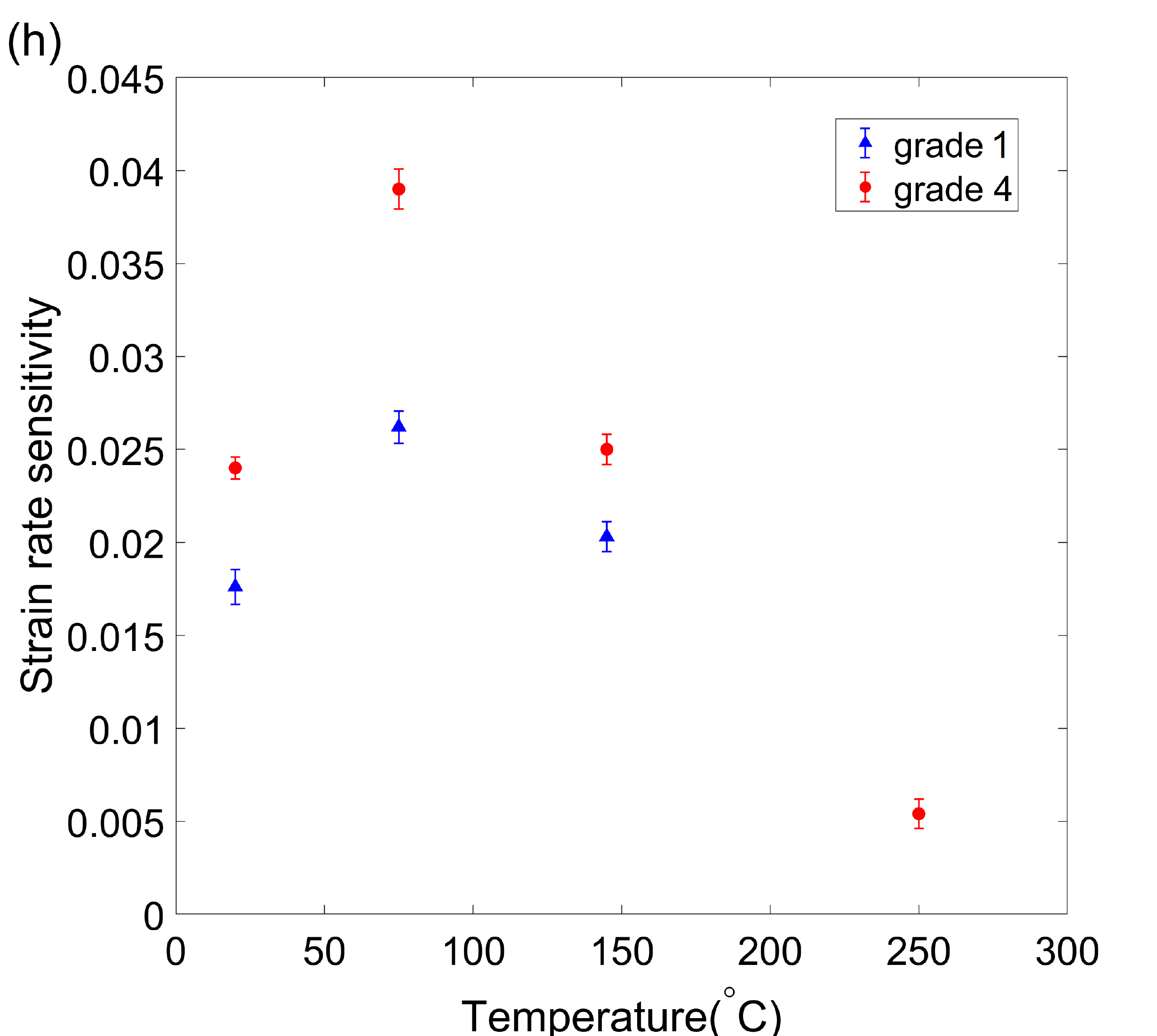}
}
\caption{Strain rate sensitivity variation against c-axis declination angle of CP-Ti grade 4 at (a) room temperature; (b) 75~$^{\circ}$C; (c) 145~$^{\circ}$C; (d) 250~$^{\circ}$C; and CP-Ti grade 1 at (e) room temperature; (f) 75~$^{\circ}$C; (g) 145~$^{\circ}$C and (h) Averaged SRS values of the 21 plane families at different temperatures.}
\label{fig.11}
\end{figure}
\begin{figure}[h!]
\centering
\subfigure{
\includegraphics[width=.5\textwidth]{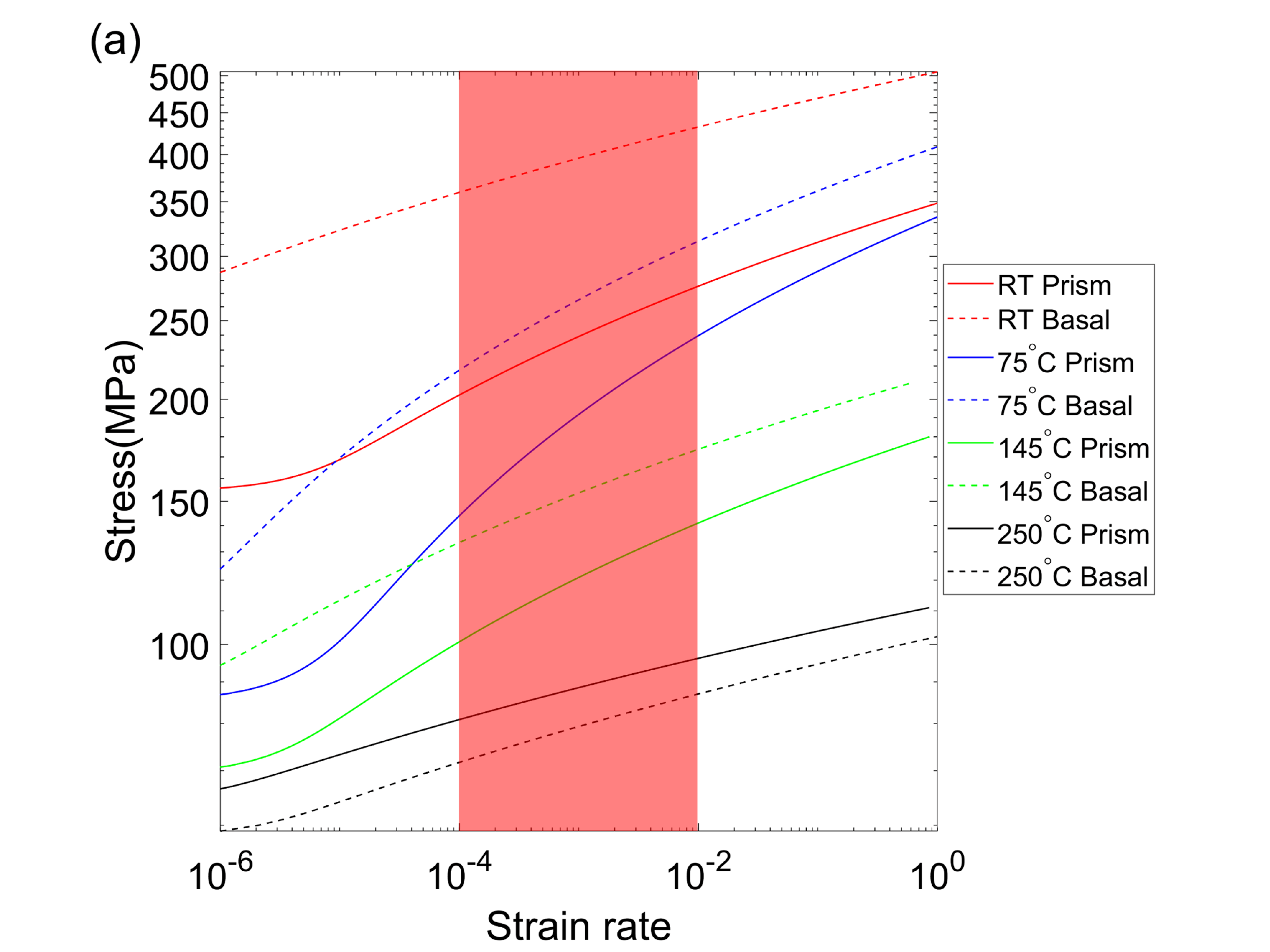}
}
\subfigure{
\includegraphics[width=.415\textwidth]{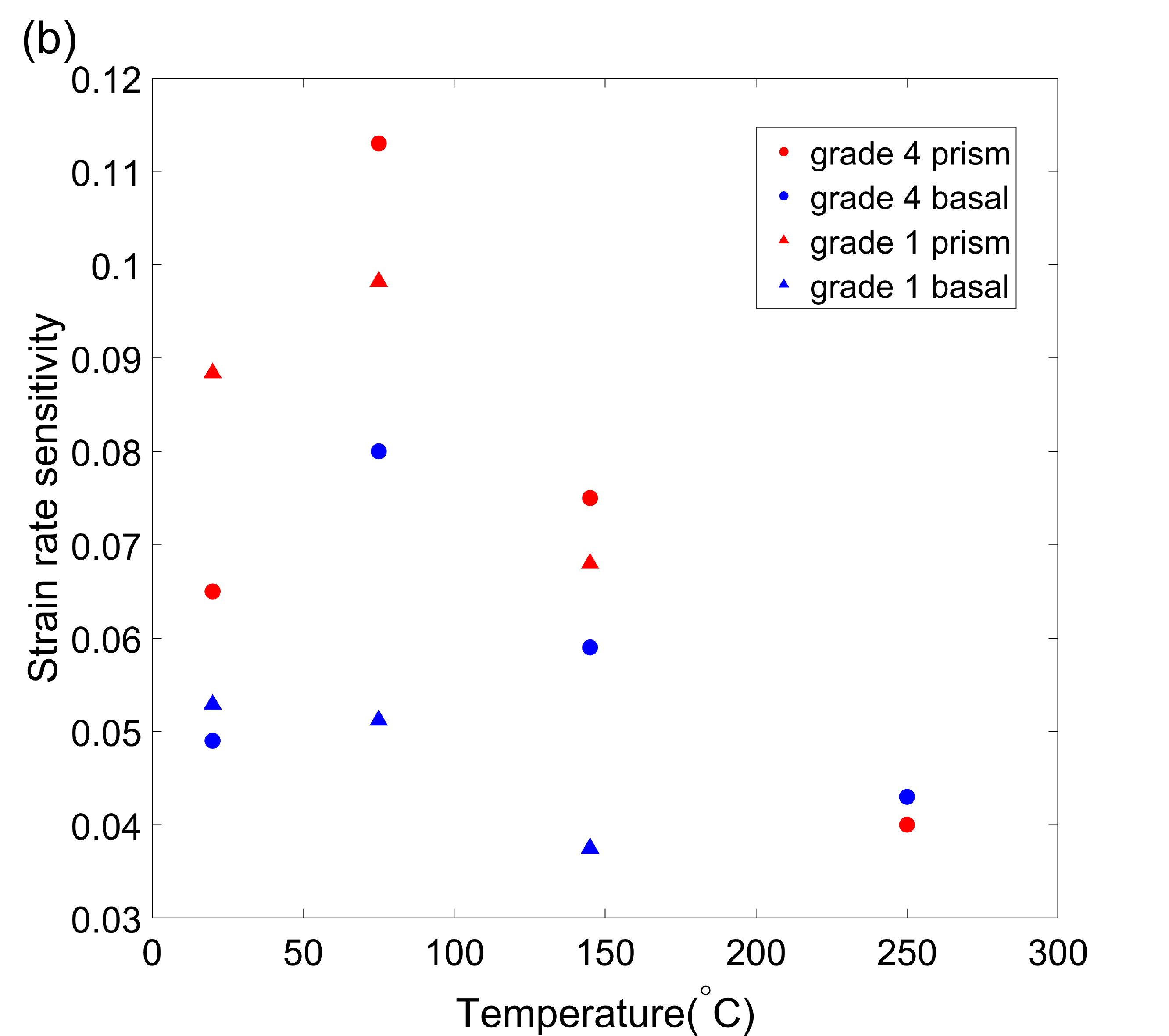}
}
\caption{(a) Stress vs. strain rate plot over strain rate range between $10^{-6}$ to $10^0~\text{s}^{-1}$ for prism and basal slips in CP-Ti grade 4 at 4 different temperatures (red shadow shows the typical strain rate range for Ti in service); (b) Variation of SRS values with temperatures for prism and basal slips. }
\label{fig.12}
\end{figure}

Similarly, the SRS can be calculated for each different slip system using the slip law (recall Eq.~\ref{eq6} in section 2.3). Inserting the slip parameters obtained from the simulation (section 3.3), we can plot the stress vs shear strain rate over a strain rate range of $10^{-6}$ to $10^0$~$\text{s}^{-1}$ (see Fig.~\ref{fig.12}(a)). As the stress vs. shear strain rate curves are not linear and the gradient is not constant at different shear strain rate range, showing the SRS is different for a different strain rate range. Therefore, a narrower strain range ($10^{-4}$ to $10^{-2}$~$\text{s}^{-1}$) was selected for reflecting the Ti aeroengine components in service and enables a better comparison to other values reported~\cite{JUN2016NANO,JUN2016,ZHANG2016Intrinsic}. Grade 4 has a higher SRS over grade 1 for both prism and basal slips. Prism slip has a higher SRS than that of basal slip in both two CP-Ti alloys, which agrees with the work done by Jun \emph{et al}~\cite{JUN2016}, where micropillars were made from the $\alpha$ phase in Ti6242 alloy and the SRS values were determined using two different methods: a constant strain rate method (CRSM) and a conventional stress relaxation method (SRM). For both methods, SRS for prism slip is more than two times the values of basal slip. Correlating to the variation of SRS against c-axis declination angle, it is believed that the difference of SRS in different slip systems is the origin of this orientation-sensitive SRS. 

The temperature dependence of SRS in basal and prism slips follow the similar trend as the averaged SRS and relaxed stress, where a peak value can be predicted between 75~$^{\circ}$C and 145~$^{\circ}$C. The peak temperature agrees with the worst-case temperature (90~$^{\circ}$C to 120~$^{\circ}$C) for dwell debit~\cite{ZHANG2015,ZHENG2017Mechanistic}. Therefore, combining the evidence collected in this work, the physical phenomenon correlating to strain rate sensitivity can be interpreted in the context of cold creep. During the same period of creep, higher strain rate sensitivity leads to greater accumulation of plastic strain. Strain rate sensitivity of each slip system is positively correlated to the activity of the slip system during stress dwell. For the lower oxygen content the CRSS of prism slip always remains significantly smaller than that for basal slip, while at higher oxygen content the slip strengths become more comparable.  Prism slip still remains the more likely contributor to load shedding and cold dwell fatigue issues due to its greater strain rate sensitivity compared to basal slip.

\section{Summary}
An in-situ X-ray diffraction method was used to measure the lattice strain evolution in two commercially pure Ti alloys (grade 1 and grade 4) with different oxygen contents at four temperatures (room temperature, 75~$^{\circ}$C, 145~$^{\circ}$C and 250~$^{\circ}$C). The samples were deformed in tension to a small plastic strain, followed by a five-minute strain hold to allow stress relaxation. The results were interpreted using a crystal plasticity finite element model. The lattice strain evolution for 21 crystal plane families during macroscopic stress relaxation was captured. Combining with the crystal plasticity finite element analysis, key parameters in the slip rule for basal and prismatic slip systems were isolated and determined. The study has the following conclusions:
\begin{enumerate}
  \item At the macroscopic level, both the activation volume ($\Delta{V}$) and the thermal activation energy ($\Delta{F}$) were found to be higher in low oxygen content CP-Ti (grade 1) and were both observed to increases in response to the increasing temperature. However, the strength decreased significantly with increasing temperature. The magnitude of stress relaxation was found to be largest at 75~$^{\circ}$C for both grade 1 and grade 4. 
  \item The CRSS and thermal activation energy were found to be higher for basal slip over prism slip except at 250~$^{\circ}$C where the trend is reversed. Due to the greater strengthening effect of oxygen on prism slip the ratio of CRSS values for grade 4 decreases significantly with temperature, while it does not for grade 1 (it may even increase slightly).   
  \item With increasing temperature, the activation energy barrier ($\Delta{F}$) increases, but as the mechanical energy required for slip reduces as the CRSS values decrease so that thermal energy becomes a greater contributor to dislocation glide processes. 
  \item Strain rate sensitivity was determined from lattice strain relaxation in 21 plane families. It shows strong orientation dependence, where grains in ‘soft’ orientation have a higher SRS than that of grains in ‘hard’ orientation. The highest average SRS among 21 plane families was found at 75~$^{\circ}$C for both two CP-Ti alloys and grade 4 has higher SRS over grade 1.
  \item The SRS for prism slip was found to be higher than that of basal slip. The temperature dependence of SRS for both slip systems follow similar trends as the average SRS and macroscopic relaxed stress, indicating that a higher SRS results in more plastic strain accumulation during creep. 
\end{enumerate}

\section*{Data Availability}
The synchrotron diffraction patterns and mechanical test data recorded during this experiment will be made openly available on the website https://zenodo.org/.

\section*{Author Contributions}
\textbf{Yi Xiong}: Data Curation, Formal Analysis, Methodology, Investigation, Software, Validation, Visualisation, Writing – Original Draft, Writing – Review \& Editing

\textbf{Phani Karamched}: Data Curation, Formal Analysis, Investigation, Methodology, Software, Supervision, Validation, Writing – Review \& Editing

\textbf{Chi-Toan Nguyen}: Investigation, Methodology, Validation, Writing – Review \& Editing

\textbf{David M Collins}: Data Curation, Formal Analysis, Investigation, Methodology, Software, Writing – Review \& Editing

\textbf{Nicolò Grilli}: Methodology, Software, Writing – Review \& Editing

\textbf{Christopher M Magazzeni}: Investigation, Writing – Review \& Editing

\textbf{Edmund Tarleton}: Data Curation, Formal Analysis, Methodology, Software, Supervision, Validation, Writing – Review \& Editing

\textbf{Angus J Wilkinson}:   Conceptualization, Funding Acquisition, Investigation, Project Administration, Supervision, Visualisation, Writing – Review \& Editing

\section*{Acknowledgements}
The authors acknowledge funding from the EPSRC through the HexMat programme grant (EP/K034332/1) and the Diamond Light Source for beam time under experiment EE17222. We are grateful for use of characterisation facilities within the David Cockayne Centre for Electron Microscopy, Department of Materials, University of Oxford, which has benefitted from financial support provided by the Henry Royce Institute (Grant ref EP/R010145/1). YX expresses gratitude to the financial support of China Scholarship Council (CSC) and ET acknowledges EPSRC for support through Fellowship grant (EP/N007239/1). We would like to thank Dr.Thomas Connolley, Dr.Robert Atwood and Dr.Stefan Michalik for their friendly and patient help at the beamline I12.

\section*{Appendix}

\renewcommand{\thefigure}{A\arabic{figure}}
\setcounter{figure}{0}

\begin{figure}[h!]
\centering
\subfigure{
\includegraphics[width=.47\textwidth]{Figures/FigA1a.pdf}
}
\subfigure{
\includegraphics[width=.47\textwidth]{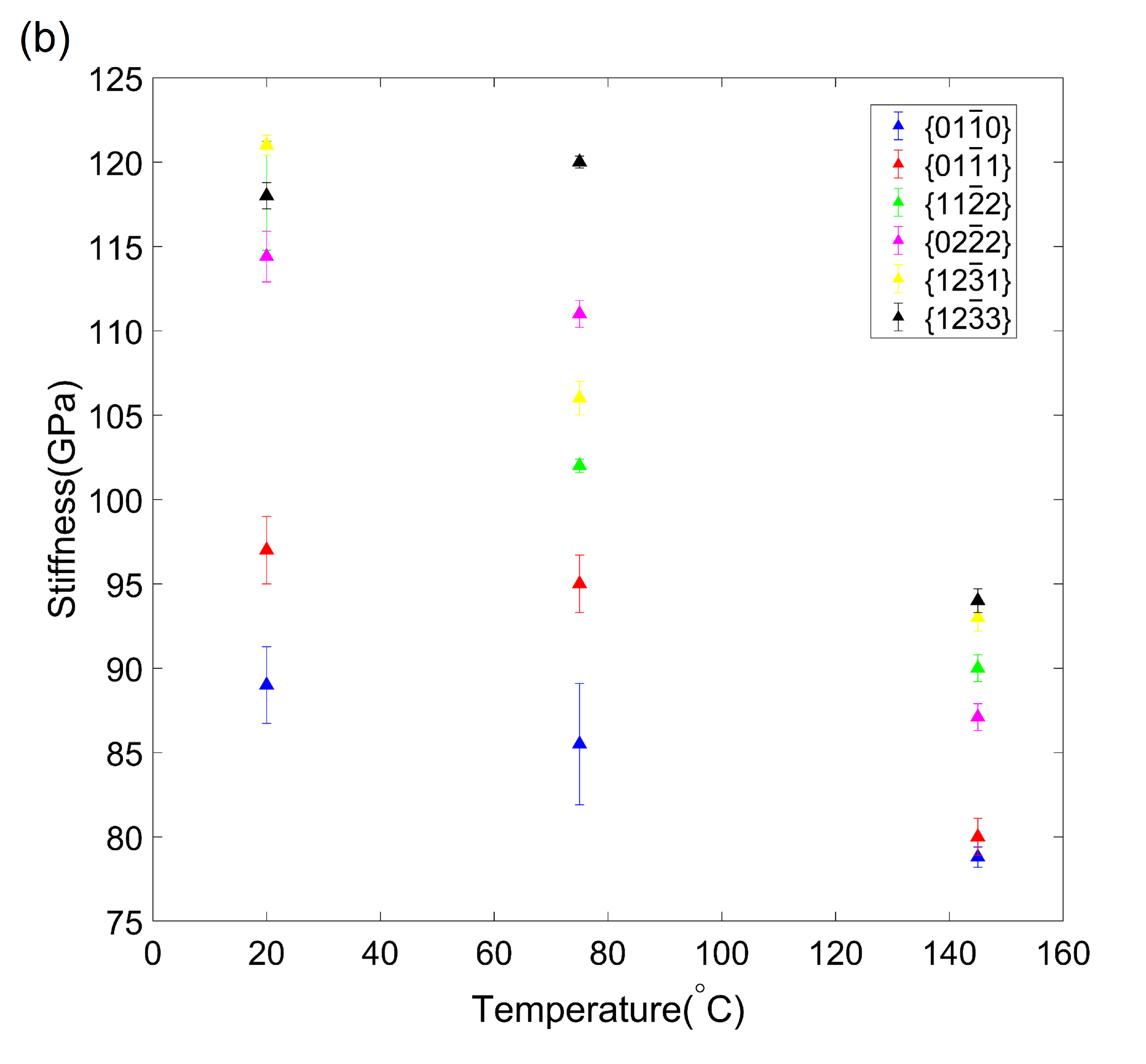}
}
\caption{(a) Thermal expansion strain in the $a$ and $c$ directions in CP-Ti grade 1 samples; (b) Variation of the stiffness of 6 plane families vs. temperature in CP-Ti grade 1. (In addition to Fig.~\ref{fig.5}) }
\label{figA1}
\end{figure}
\clearpage
\begin{figure}[t!]
\centering
{
\includegraphics[width=.475\textwidth]{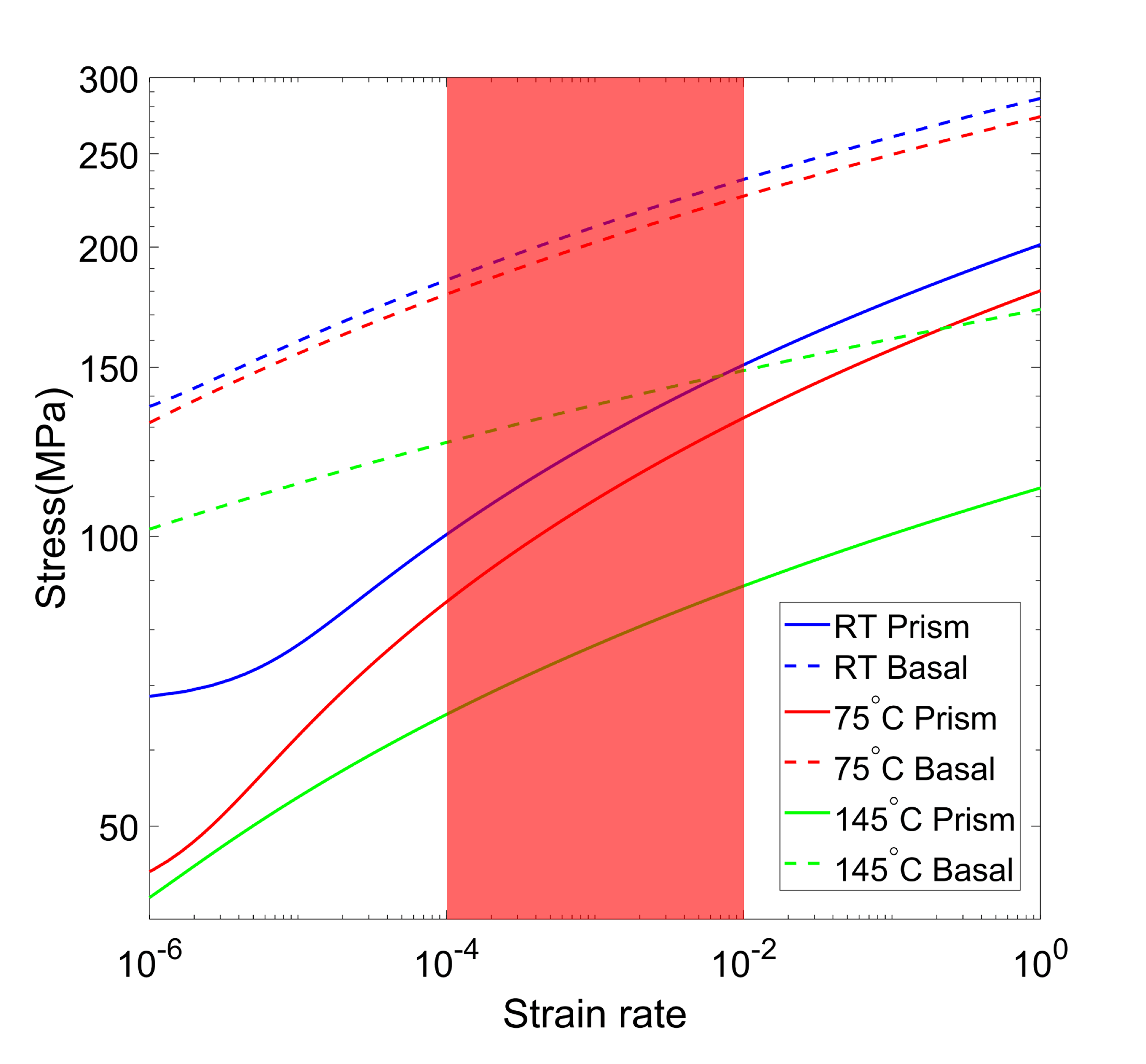}
}
\caption{Stress vs. strain rate plot over strain rate range between $10^{-6}$ to $10^0~\text{s}^{-1}$ for prism and basal slips in CP-Ti grade 1 at 3 different temperatures, the red shadow shows the typical strain rate range for Ti in service. (In addition to Fig.~\ref{fig.12}) }
\label{figA2}
\end{figure}

\bibliographystyle{elsarticle-num}
\bibliography{Effect_of_temperature_and_oxygen}

\end{document}